# Consensus for Compressed Static Functions

**Dominik Rosch**
Karlsruhe Institute of Technology, Germany

**Jonatan Ziegler**
Karlsruhe Institute of Technology, Germany

## Abstract

The Consensus technique marked a breakthrough in the construction of minimal perfect hash functions (MPHFs), reaching a linear tradeoff between construction time and space overhead relative to the optimum. Consensus provides a clever scheme to search for and encode seeds of *tasks* in random data structures. In this paper, we apply Consensus to the related field of compressed static functions (CSFs). These data structures store a function $f : S \to \Sigma$ such that querying a key $x \in S$ returns $f(x)$ and querying $x \notin S$ returns an arbitrary value. Because of this freedom, CSFs do not need to store the keys $S$ and only need space close to the zeroth-order empirical entropy of the multiset of values. Often, some values are much more common than others. In these cases, CSFs can use less space than their non-compressed counterparts, which are optimized for a uniform value distribution. CSFs are a useful building block, for example in database design and bioinformatics.
We introduce Consensus-CSF, which can reach arbitrarily close to the empirical entropy $nH_0$, with a construction time of $n \exp\left(\tilde{\mathcal{O}}\left(\sqrt{1/\delta}\right)\right)$ for space usage of $nH_0(1+\delta)$ when assuming some parameters of the value distribution to be constants. This tradeoff beats previously implemented approaches that can only reach some fixed threshold above the entropy lower bound. We enable Consensus in the setting of CSFs, which is less structured than MPHFs, with the introduction of *task insertions*. Our approach randomly distributes the keys into one-bit Consensus tasks and then strategically *inserts* additional tasks in places where the construction would get stuck otherwise. We provide an implemented version of our algorithm which reaches the same order of magnitude in space overhead as competitors but is not competitive in practice. Beyond these results, we present a new way to think and reason about Consensus, which may also be applied to other problems, like minimal perfect hash functions with fast query times.

**2012 ACM Subject Classification** Theory of computation → Randomness, geometry and discrete structures; Theory of computation → Data compression; Theory of computation → Data structures design and analysis

**Acknowledgements** This paper is the result of a student research project supervised by Stefan Hermann, Ragnar Groot Koerkamp, and Stefan Walzer. We used the aid of LLMs for literature research, Chernoff bounds, scripting of the experiments and proofreading.

## 1 Introduction

It has been known for a long time that entropy is a measure of the minimal space needed for storing data [23]. Yet for many problems, no practical data structures reach this limit.

**Static Functions.** One such problem is *static functions*, also known as *retrieval data structures*, which are a useful building block for database systems [3, 20] and used in bioinformatics [24]. A static function is defined as follows:

▶ **Definition 1** (Static Function). *Given a finite set of keys $S \subseteq U$ of some universe, a finite value set $\Sigma$, and a function $f : S \to \Sigma$ that assigns each key a value. A* static function *(*retrieval data structure*) is a data structure that encodes a function $h_f : U \to \Sigma$ where $h_f|_S = f$, i.e., $\forall x \in S : h_f(x) = f(x)$.*

Note that the provided function $h_f$ can return arbitrary values outside of $S$. This allows a static functions data structure's space usage to only depend on the values, not on the keys.

Good solutions reach construction time linear in $n := |S|$ and query time independent of $n$. Regarding space usage, there already exist succinct solutions when assuming values are uniformly distributed [6], reaching $n \log_2|\Sigma|$ bits of space usage. Thus, in the following, we consider a more generic setting where values $\sigma \in \Sigma$ appear with different frequencies, but we do not consider any correlation between keys and values. This shifts the space goal down to the *zeroth order empirical entropy* $nH_0$ of the values $(f(x))_{x\in S}$ where $H_0 := \sum_{\sigma\in\Sigma} \rho_\sigma \log_{1/2} \rho_\sigma$ and $\rho_\sigma := \left|f^{-1}(\sigma)\right|/n$ is the frequency of value $\sigma \in \Sigma$. Data structures aiming for this space goal are called *compressed* static functions (CSFs).

Every known CSF data structure cannot reach arbitrarily close to this entropy [3, 6–8, 11, 12, 19] or is impractical and not implemented [1]. In this paper, we close this gap and provide a CSF data structure that fulfills both criteria—an arbitrarily small space overhead in theory together with a practical implementation. Our idea is to apply Consensus [17] to CSF construction. Recently, the Consensus technique made getting close to entropy practical in the related field of *minimal perfect hash functions* (MPHFs). An MPHF is essentially a CSF with $\Sigma = \{0, ..., |S| - 1\}$ and $f$ being a bijection where any concrete mapping from keys to values is considered equally good.

Consensus solves the *seed search and encoding problem* (SSEP) [17], that is, given a sequence of Bernoulli processes, called *tasks*, construct a bitstring that encodes a successful index—a *seed*—for each Bernoulli process. Consensus solves the SSEP by giving tasks a fixed number of tries, and if they fail, going back to previous tasks, which yields independent chances for future tasks. We explain Consensus in more detail in Section 2.1.

**Consensus-CSF.** Mapping CSF construction to the SSEP is not as straightforward as it is for MPHFs, as CSFs do not require an ordering of keys as MPHFs do. We reduce CSF construction to an SSEP as follows: Ideally, each key $x \in S$ would correspond to a task, which is successful if, by chance, we select its value $f(x)$ out of all values $\Sigma$. We use a seeded biased hash function with range $\Sigma$ for this selection. The probability that we select a value $\alpha \in \Sigma$ is equal to its frequency $\rho_\alpha$. Yet, this approach requires mapping keys to tasks to begin with, which would essentially be an MPHF with all its space usage wasted. Thus, we use a weaker task assignment scheme. We fix each task at two tries (*one-bit tasks*) and allow some tasks to receive more or fewer keys. This is acceptable as Consensus can handle some imbalance. We first partition keys uniformly into *groups*, and then inside a group to a fixed number of tasks. Initially, groups are a bit *overloaded* compared to how much information their tasks can encode. If a group receives too many keys by chance, we mark this group and insert additional tasks. More tasks make each task inside this group receive fewer keys in expectation. Sometimes, this is still not enough, or the mapping of keys to tasks inside the group is unfortunate. Then, we bump all this group's keys to a less ambitious fallback. CSFs pose another problem. As some values are less common than others, for some keys it is harder to find their value. This shifts the problem from balancing the number of keys inside each group and task to balancing their information content, their *weight*. Our balancing scheme of task insertion can be adapted to be weight-based easily.

### 1.1 Notation and Theoretical Model

We use $\log := \log_2$, and $\log_{1/2} x = \log \frac{1}{x} = -\log x$ to avoid fractions and negatives. Additionally, $0 \in \mathbb{N}$, $[n] = \{0, ..., n-1\}$ for some $n \in \mathbb{N}$ and all indices start from 0.

**Hashing Assumption.** We assume that we have access to a uniform seeded hash function $\xi_s : U \to [0,1]$, $s \in \mathbb{N}$ available, where $\xi_s(x) \overset{\text{iid.}}{\sim} U([0,1])$ for all $x \in U, s \in \mathbb{N}$. We assume evaluating it takes constant time, and we need no space to store it.

### 1.2 Results & Contribution

We demonstrate that the Consensus technique can be applied to more unpredictable settings than RecSplit-based MPHF construction. This becomes possible through the novel technique of strategically *inserting tasks* to prevent excessive backtracking. We show that with these insertions, the task difficulty can be effectively contained inside an interval. For this new balancing approach, we provide a new ground-up analysis of Consensus' performance. Using task insertions, we develop a novel compressed static function data structure we call Consensus-CSF. It has constant query time and a space-usage–construction-time tradeoff of $nH_0(1+\delta)$ to $n \exp\left(\tilde{\mathcal{O}}\left(\sqrt{1/\delta}\right)\right)$ when assuming some properties of the value distribution to be constants, see Theorem 2. Additionally, we provide an implementation [21] of our data structure and compare it to previous approaches.

In the following sections, we first briefly give an overview of related work. Then, we introduce and explain our algorithm in full detail. Afterward, in the main part of our paper, we analyze the performance of our algorithm. Finally, we provide results of our implementation. Appendix A gives an overview of the variables used, and Appendix B provides some additional proofs.

## 2 Related Work

In this section, we give an overview of the existing approaches for constructing a CSF.

**Belazzougui and Venturini.** Belazzougui and Venturini [1] provide a CSF which reaches space usage arbitrarily close to $nH_0$ for large enough inputs. However, their CSF is purely theoretical and has never been implemented. Their approach utilizes a 2D table filled with random values according to their frequency in the input. Keys are partitioned into buckets but also mapped injectively to columns of a table. These two functions use a theoretical MPHF from Hagerup and Tholey [10]. For each bucket, Belazzougui and Venturini select a row of the table such that every key attains its designated value in the column it was mapped to. The main space usage stems from storing the indices of the selected row for each bucket.

**Linear equations.** A more practical CSF was introduced by Hreinsson, Krøyer, and Pagh [12]. Their approach has a linear overhead over the optimal space usage, but has been implemented in various forms and applied [3, 7, 8, 12, 19, 24]. At its core, this approach utilizes a prefix-free code and a bit array. To query a key, it determines $k$ positions in the array with hash functions. Then, it combines the values at these positions with XOR. This is repeated for the next position to the right and so on, until a complete codeword is read. This yields the correct codewords if the bit array satisfies a system of linear equations in $\mathbb{F}_2$. For this system to be solvable, the satisfiability threshold of $k$-XORSAT gives a bound on the required array size, which dictates the space overhead above $nH_0$. One drawback of this approach is its dependence on prefix-free codes. Even the best possible such code can have up to one bit overhead. For example, consider a very common value that can carry arbitrarily small information. Then, a prefix-free code cannot do better than encode that common value with one bit. This can to some extent be mitigated by adding an approximate membership query data structure to filter out the most common value. One representative of this family is Caramel [3], against which we compare our CSF later on.

**Membership-based.** Hermann et al. [11] provide another approach to building CSFs as a base case when introducing learned static functions. Their approach is to build a coding tree for the given distribution of values, e.g., Huffman codes [14]. When decoding a value, each inner node in the coding tree corresponds to a decision: descending left or right. The probability of taking a decision depends on the node in the tree—on how many keys shall be mapped to values lying below the left or right child of this node. Hermann et al. store and encode these decisions for all keys using a weighted relative membership data structure, based on a weighted filter data structure, based on a variable-length retrieval data structure, based on BuRR [5, 6], a non-compressed retrieval data structure. This extends the *filter trick* mentioned in linear equation–based approaches, applying it at every node, not only once at the root of the coding tree. BuRR itself also uses a system of linear equations in $\mathbb{F}_2$ to encode the mapping of keys to values. Should this system be unsolvable, BuRR bumps some of its keys to a fallback BuRR recursively. We refer to this approach for building a CSF based on BuRR as *BuRR-CSF* in the following. They also only reach a fixed linear overhead.

## 2.1 Consensus

As it forms the basis of our CSF, we now introduce the Consensus technique in more detail. In the field of randomized data structures, a common approach is to try out a random data structure—identified by a seed—until one satisfying a condition is found. Finding such a data structure is particularly efficient when splitting it into smaller parts that can be searched independently. This motivates the seed search and encoding problem (SSEP) [17]: Provided a Bernoulli process $\vec{B}^{(i)}$ for each *task* $i \in [N]$, the SSEP demands a bit string that provides an encoding for a sequence of seeds $(s_i)_{i\in[N]}$ such that $B^{(i)}_{s_i} = 1$, that is, the $s_i$-th try in the $i$-th Bernoulli process is a success.

Consensus solves the SSEP better than naive approaches with a bounded brute-force search across all tasks with backtracking, together with a clever encoding scheme making future tasks dependent on previous ones. Figure 1a shows how a search tree for Consensus' brute-force search might look. Each numbered layer represents one task with its Bernoulli process. A dashed edge marks a failed try for a task. Consensus essentially performs a DFS on this graph to find an unbroken path through all layers. The root layer provides an infinite number of retries to guarantee success—its tries are stored in a *root seed.*

Note that it is sufficient to only store the seed in the last layer, as from it the entire path becomes clear. In practice, $k_i$ are chosen as powers of two. Then, all the seeds are just prefixes of the base two bit string representation of the last seed. Furthermore, instead of a linearly sized prefix being considered as this root seed, in practice it is sufficient to only consider the last word of it (then of course the exact successful path changes, yet one for this seed scheme can still be found). Figure 1b shows how this seed representation looks.

When allowing for enough tries in each task, Consensus only needs to backtrack an expected constant number of times per task. The construction-time–space tradeoff can be tuned by changing the number of tries for each task.

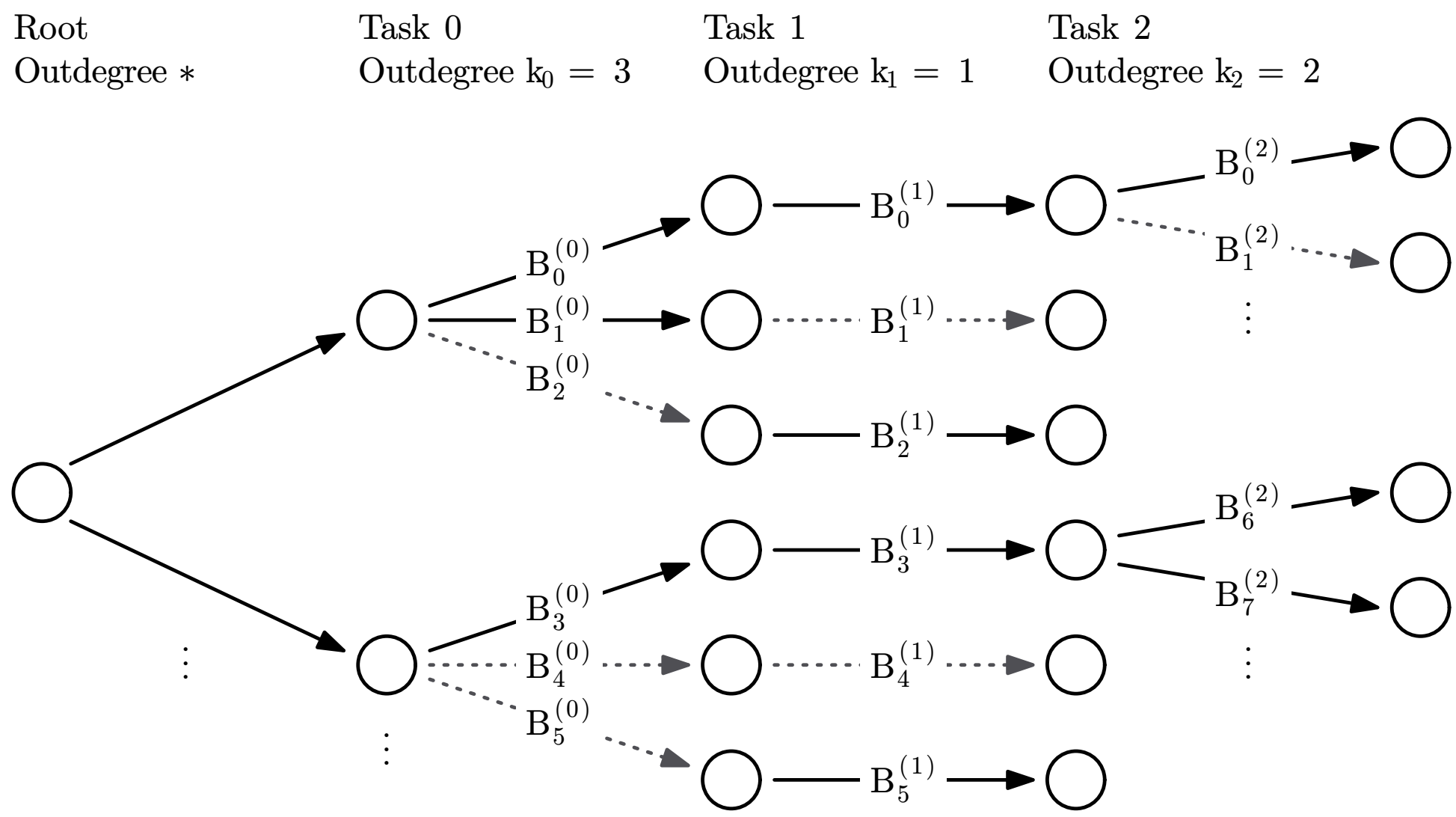

(a) CONSENSUS' search tree. The root layer contains the root seed; each of the following layers corresponds to one task—one Bernoulli process.

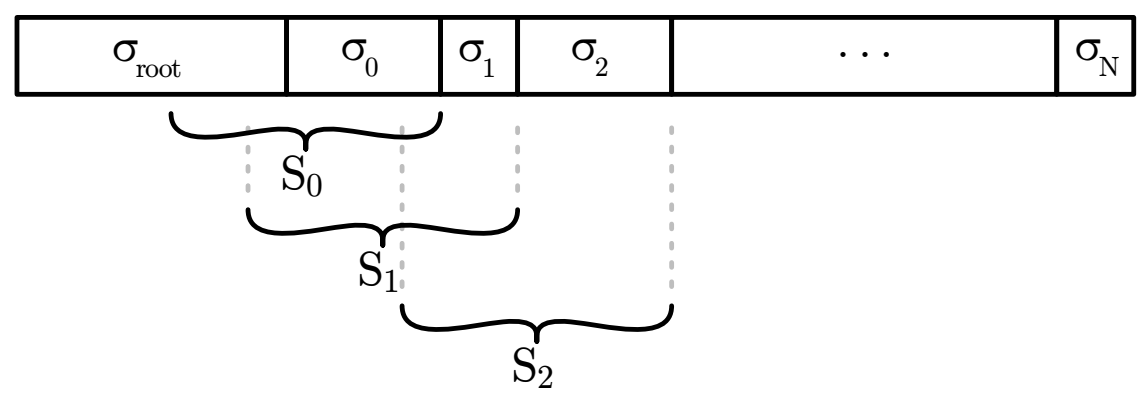

(b) How CONSENSUS encodes the seeds. $\sigma_i$ is the local index of the out-edge chosen in task $i$.

**Figure 1** Depiction of CONSENSUS' search tree (1a) and seed representation (1b), adapted from [17].

For the analysis of the number of seeds that need to be tried for each task, it is useful to consider the probability $q_t$ that CONSENSUS will not backtrack from task $t$, i.e., all remaining tasks will be successful. If the probability that a seed is successful for task $t$ is $p_t$, we get the recurrence

$$q_t = 1 - (1 - p_t q_{t+1})^{k_t}$$

where $k_t$ is the number of seeds that can be tried in task $t$. This equation simply states that CONSENSUS will backtrack out of task $t$ if during $k_t$ independent tries either task $t$ is not successful ($p_t$) or task $t+1$ will backtrack again ($q_{t+1}$). This recurrence is rooted in $q_{\text{end}} = 1$, where CONSENSUS will never backtrack as there are no more tasks—the search is complete. Task $t$ then needs $\text{Geo}(p_t q_{t+1})$ seeds tested. Sometimes, with probability $1 - q_0$, CONSENSUS will backtrack from the first task. Then, we need a variable-length root seed at the start of the bit vector to just retry the entire search. This root seed is $\text{Geo}(q_0)$ distributed.

When applying CONSENSUS to CSF construction, we will always choose $k_t = 2$ as we store only one bit per task, yet $p_t$ is a random variable.

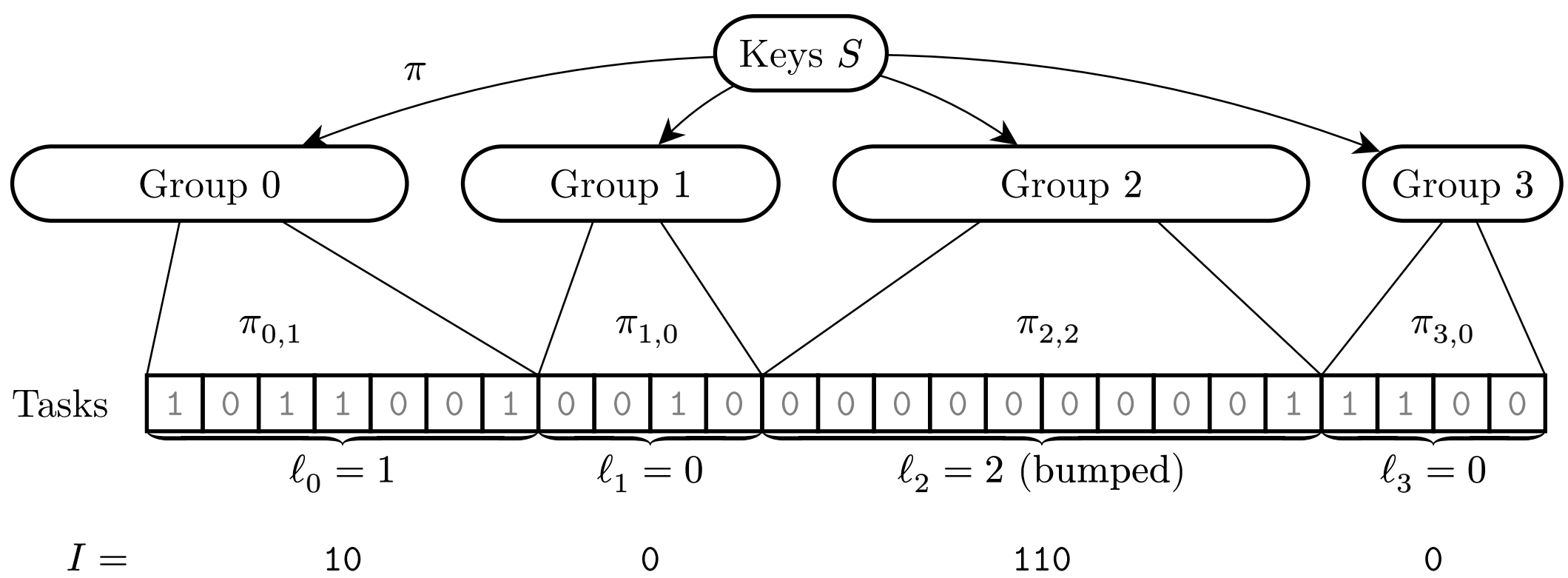

**Figure 2** Mapping of keys to tasks. First, keys are uniformly and randomly mapped to groups using $\pi$. Then, depending on a group's difficulty, the number of insertions $\ell_i$ is chosen. Afterwards, keys are mapped uniformly to an interval of $b + \ell_i\beta$ tasks using $\pi_{i,\ell_i}$. The bits of the insertion vector $I$ are shown at the bottom.

## 3 Consensus Retrieval Algorithm

In this section, we describe the algorithm in detail and introduce necessary notation.

**Hashing retrieval values.** Like the problem of minimal perfect hash functions where Consensus originated, we can solve the retrieval problem by searching for seeds for the keys. These seeds $s \in \mathbb{N}$ parameterize hash functions $h_s : U \to \Sigma$ from the universe of keys to the values where $h_s(x)$ is distributed according to the empirical value distribution, i.e., $\Pr[h_s(x) = \alpha] = \rho_\alpha = |\{y \in S : f(y) = \alpha\}|/n$ for $x \in U$ and $\alpha \in \Sigma$. Note that for finite $\Sigma$ we can create such hash functions with constant evaluation time from a family of seeded uniform hash functions $(\xi_s : U \to [0, 1])_{s\in\mathbb{N}}$ using an alias table [27].

**Retrieval with Consensus.** The high level idea is that every key $x \in S$ is (non-injectively) assigned to a task $t$ in the Consensus bit vector $C$ such that a $w$ bit word up to $t$ in $C$ is a seed $s$ such that $h_s(x) = f(x)$. A naive approach is to uniformly partition the keys to one-bit tasks which correspond to the positions in the Consensus bit vector, and then use the Consensus backtracking algorithm to fill the bit vector. The problem with this approach is that with high probability tasks will be very difficult because they or the tasks after them received a lot of *heavy* keys—keys with rare values. This leads to a lot of backtracking, and makes the construction time impractical.

**Task partitioning.** Instead, we partition the keys to tasks in two steps, see Figure 2: First, a key $x$ is mapped uniformly to a group $i = \pi(x) \in \left[\frac{n}{\lambda}\right]$ of consecutive one-bit tasks in $C$ using a hash function, with $\lambda$ being the expected number of keys inside a group. Then, inside the group, $x$ is uniformly mapped to a task $j = \pi_{i,\ell_i}(x) \in [b + \ell_i\beta]$ using a hash function. Initially, a group starts out with $b \in \mathbb{N}$ one-bit tasks and $\ell_i = 0$. If the difficulty of this group is not *good* (formally defined in Definition 3 later), e.g., there is a too difficult task, we *insert* $\beta$ additional one-bit tasks. This also reshuffles all the keys in the group and leads to a more balanced group in expectation.

**Bumping.** In the unlikely case that the difficulties are still bad, we insert a second time to mark it as a bad group and bump all associated keys to a fallback data structure. Note that these groups sill take up space. Thus, the number of possible insertions is $\ell_i \in \{0, 1, 2\}$.

**Insertion vector.** We save the number of insertions $\ell_i$ for each group using a unary encoding $1^{\ell_i}$ in the *insertion vector* $I$. Here, we separate the groups by a zero bit. We

**Algorithm 1** Construction of solution to the problem using CONSENSUS. A concrete definition of the *good* predicate used is given in Algorithm 3 which depends on notation only defined later in Section 4 Analysis.

```
    function CSF::construct(f : S → Σ, b, β, λ, good)
 1    I := ⟨⟩; S_F := ∅
 2    partition S into S_i := {x ∈ S : π(x) = i} for i ∈ [n/λ]
 3    for all groups i = n/λ − 1, ..., 0 do
 4        for ℓ_i ∈ {0,1} do
 5            partition S_i into tasks S_{i,j} := {x ∈ S_i | π_{i,ℓ_i}(x) = j} for j ∈ [b + ℓ_i β]
 6            if good(i, ℓ_i, (S_{î,j})_{(î,j)≥(i,0)}) then
 7                I := 1^{ℓ_i} 0 ∘ I
 8                continue with next group
 9        S_F := S_F ∪ S_i                       // bad group ⇒ bump to fallback
10        ℓ_i := 2; I := 1^{ℓ_i} 0 ∘ I
11    h_(·) := AliasTableHasher::construct({(α, ρ_α) : α ∈ Σ})
12    C := CONSENSUS::construct(tasks := ({∀x ∈ S_{i,j} : h_s(x) = f(x)})_{s∈ℕ} for i∈[n/λ], j∈[b+ℓ_i β])
13    F := Fallback::construct(S_F)
14    I := Select::construct(I)
15    return (insertion vector I, CONSENSUS vector C, fallback F, alias table hasher h_(·))
```

augment the insertion vector with a select data structure so that we can find the position of the $i$-th zero in constant time and thus can calculate where the $(i+1)$-th group starts.

In the end, we store only the insertion vector $I$, the *CONSENSUS vector* $C$, the fallback $F$, and the alias table.

**Construction.** Concretely, the construction happens with Algorithm 1 in two *phases*. In the first *insertion phase*, we look at each group right-to-left and check if inserting zero times or one time ($\ell_i \in \{0,1\}$) leads to a good group, or if we need to bump its keys into the fallback ($\ell_i = 2$). In the second *CONSENSUS phase*, the normal CONSENSUS backtracking algorithm finds a suitable seed for every task. We also construct the fallback data structure.

**Query.** We query a key $x \in U$ with Algorithm 2. First, we hash to the group index $i = \pi(x)$ and the find the position of the $(i+1)$st zero in the insertion vector $I$. From this we can find $\ell_i \in \{0,1,2\}$ as the number of contiguous ones preceding that position, and how much was inserted in previous groups, which determines the start $o$ of the group $i$.

If $\ell_i \in \{0,1\}$ we look at task $j = \pi_{i,\ell_i}(x)$ in the group and get the seed $s$ from $C$ as the $w$-bit word ending at $o + j$. Then, $h_s(x) \in \Sigma$ is the value we return, which is by construction $f(x)$ if $x \in S$. Otherwise, we query the fallback data structure $F$, which is discussed in Section 4.3.

**Utilized data structures.** For our theoretical analysis, we use a select data structure like [18:4.4.3] with linear construction time, constant queries, and linear space usage. Further, we use an alias table for $h_s$ [27], which can be constructed in $\mathcal{O}(|\Sigma|)$, queried in constant time, and uses space $\mathcal{O}(|\Sigma|)$. Finally, we use a side note from [1] as the fallback, see Section 4.3 for more details.

**Result.** We formulate the following theorem about the accomplishment of our algorithm:

**Algorithm 2** The query function.

```
function CSF::query(x ∈ U,
(insertion vector I, Consensus vector C, fallback F, alias table hasher h_(·))) → Σ
1   i := π(x)                                              // group index
2   z := I.select_0(i)                                     // index of the (i+1)st zero
3   ℓ_i := number of contiguous ones preceding position z in I   // ∈ {0,1,2}
4   if ℓ_i ∈ {0,1} then
5   |  j := π_{i,ℓ_i}(x)
6   |  o := b·i + β·(z − ℓ_i − i)                          // group start
7   |  s := C.seed_of_task(o + j)
8   |  return h_s(x)
9   else
10  |  return F.query(x)
```

▶ **Theorem 2** (Main Theorem). *Let $f : S \to \Sigma$ be an input for the retrieval problem of size $n := |S|$ (Definition 1). Let $\rho_\alpha := \left|f^{-1}(\alpha)\right|/n$ be the relative frequency for each value $\alpha \in \Sigma$. Further, let $I_{\text{min}} := \min_{\alpha\in\Sigma} \log_{½} \rho_\alpha$, $I_{\text{max}} := \max_{\alpha\in\Sigma} \log_{½} \rho_\alpha$ be the minimal/maximal Shannon information of a value according to the empirical distribution, $H_0 := \sum_{\alpha\in\Sigma} \rho_\alpha \log_{½} \rho_\alpha$.*

*There exists a $c > 0$ such that for every $\delta \in (0, 2^{-cI_{\text{max}}})$ there exists an algorithm that solves the retrieval problem with the following properties:*

- ***Expected construction time:***
$$\mathbb{E}[\text{Construction Time}] \leq n \cdot \left(2^{\mathcal{O}\left(\sqrt{1/\delta}\log 1/\delta\right)} + \mathcal{O}(H_0 + H_0/I_{\text{min}})\right)$$
- ***Expected space usage:***
$$\mathbb{E}[\text{Space}] \leq nH_0 + \delta n \cdot \mathcal{O}(H_0 + H_0/I_{\text{min}} \cdot \log I_{\text{max}}) + \mathcal{O}\left(\sqrt{1/\delta}\log 1/\delta + |\Sigma|\right)$$
- ***Query time*** $\Theta(1)$

In short, our algorithm has construction time linear in $n$, constant queries, and has a space usage to construction time tradeoff of $nH_0(1+\delta)$ to $n\exp\left(\tilde{\mathcal{O}}\left(\sqrt{1/\delta}\right)\right)$ when fixing $|\Sigma|$, $I_{\text{min}}$, and $I_{\text{max}}$. The $\tilde{\mathcal{O}}$ hides logarithmic factors.

## 4 Analysis

In the following, we prove Theorem 2. The main difficulty lies in bounding the success probabilities of the Consensus tasks for a suitable insertion strategy. To do this, we formally define a notion of *good* groups where these probabilities are bounded and show that a group is good with high probability. For these good groups, analyzing space usage and expected construction time is thus easier. For bad groups, which we bump, we show that the fallback is in our budget. Our proof is split into multiple lemmata; an overview is shown in Figure 3.

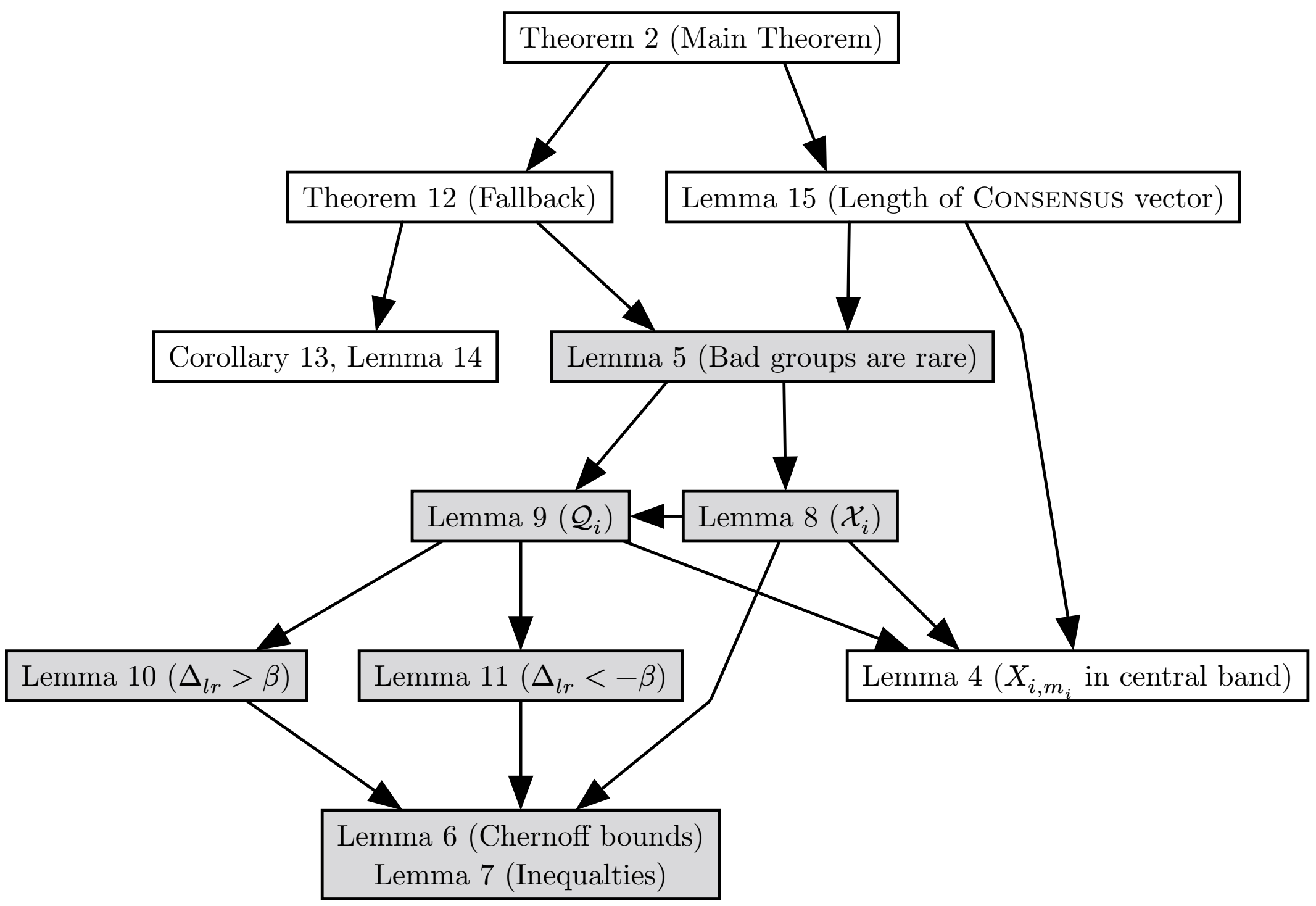

**Figure 3** Overview of lemmata and their relationships. Gray lemmata are defined and used in Section 4.2.

## 4.1 Formal Setting and Proof Idea

To make precise statements, we need to introduce some formal machinery:

**Task weights.** Algorithm 1 partitions keys into $n/\lambda$ groups of size $N_i := |S_i| \sim \text{Bin}\big(n, \frac{\lambda}{n}\big)$ ($i \in [n/\lambda]$). For a group we define its *weight* as

$$W_i := \sum_{x \in S_i} \log_{½} \rho_{f(x)} = \sum_{\alpha \in \Sigma} N_i^{(\alpha)} \log_{½} \rho_\alpha$$

where $N_i^{(\alpha)} := |\{x \in S_i : f(x) = \alpha\}| \sim \text{Bin}\big(n\rho_\alpha, \frac{\lambda}{n}\big)$ is the number of keys with value $\alpha \in \Sigma$ in the group as there are $\rho_\alpha n$ keys with value $\alpha$. Note that $N_i^{(\alpha)}$ are stochastically independent for a fixed $i$. For the expected weight of a group, we thus get

$$\mathbb{E}[W_i] = \sum_{\alpha \in \Sigma} \mathbb{E}\Big[N_i^{(\alpha)}\Big] \log_{½} \rho_\alpha = \lambda \sum_{\alpha \in \Sigma} \rho_\alpha \log_{½} \rho_\alpha = \lambda H_0.$$

For $\ell_i = \ell$ insertions, the algorithm distributes the keys of group $i$ to $m_i := b + \ell\beta$ tasks. A task $t = (i, j)$ thus receives $N_t := |S_t| \sim \text{Bin}\Big(n, \frac{\lambda}{n(b+\ell\beta)}\Big)$ keys or weight $W_t := \sum_{x \in S_t} \log_{½} \rho_{f(x)}$ as they get uniformly distributed over $b + \ell\beta$ tasks.

We index tasks of group $i$ with $t = (i, j)$ for $j \in [m_i]$ where $m_i := b + \ell_i\beta$, write $t + r := (i, j + r)$ if $t = (i, j)$, $j + r < m_i$, and define an order on these indices lexicographically.

**Task probability.** Each task $t$ has a probability

$$p_t := \Pr[\forall x \in S_t : h_s(x) = f(x)] = \prod_{x \in S_t} \rho_{f(x)} = 2^{-W_t}$$

that a seed $s$ is successful for all keys $S_t$ of that one task. The definition of weight above is motivated by $W_t = \log_{1/2} p_t$. As stated in [17], CONSENSUS' construction time depends mainly on the tail probability $q_t$ that CONSENSUS never has to visit task $t$ again. As stated in Section 2.1, for those probabilities it holds for $j \in [m_i]$

$$q_{i,j} = 1 - \left(1 - p_{i,j} q_{i,j+1}\right)^2, \quad q_{i,m_i} := q_{i+1,0}, \quad q_{n/\lambda,0} := 1.$$

**Clamping.** To aid our analysis, we consider a lower bound $\tilde{q}_t \le q_t$ which we define analogously, but with a *clamping* of $\tilde{q}_{i,m_i} := \min\{\tilde{q}_{i+1,0}, \epsilon 2^{-\beta}\}$ between the groups for easier analysis. This is in fact a lower bound since $1 - \left(1 - p_{i,j}\tilde{q}_{i,j+1}\right)^2$ is a strictly increasing function in $\tilde{q}_{i,j+1}$. Moreover, we consider this sequence in the *logarithmic domain*: $X_t := \log_{1/2} \tilde{q}_t \ge 0$. This reveals an easier to understand relationship between $q_t$, $q_{t+1}$, and the weight $W_t := \log_{1/2} p_t$ inside a group:

$$\begin{aligned} X_t &= \log_{1/2} \tilde{q}_t = \log_{1/2}\left(1 - (1 - p_t\tilde{q}_{t+1})^2\right) = \log_{1/2}\left(2p_t\tilde{q}_{t+1} - (p_t\tilde{q}_{t+1})^2\right) \\ &= \log_{1/2}((2p_t\tilde{q}_{t+1}(1 - p_t\tilde{q}_{t+1}/2)) = X_{t+1} + W_t - 1 + \underbrace{\log_{1/2}(1 - p_t\tilde{q}_{t+1}/2)}_{=:\text{loss}_t} \end{aligned}$$

**Loss.** Ignoring the *loss* term, $X_t$ form a random walk, with $t$ from large to small. Each step we gain one bit of information but have to pay $W_t$ bits. The loss term is only significant if $X_{t+1}$ is small: As $\log_{1/2}(1 - \frac{x}{2}) \le x \quad (x \in [0,1])$ we have $\text{loss}_t \le p_t\tilde{q}_{t+1} \le 2^{-X_{t+1}}$. In particular, if $X_{t+1} \ge a$ we have $\text{loss}_t \le \epsilon := 2^{-a}$. This perspective is beneficial as, when ignoring the loss, the exact position a key lands over some interval of tasks $[l, r) = \{l, l+1, ..., r-1\}$ of length $\omega$ is irrelevant for the change in $X_t$ across the tasks. This can be seen from

$$\Delta_{lr} := X_l - X_r = W_{lr} - \omega + \sum_{t=l}^{r-1} \text{loss}_t$$

where $W_{lr} := \sum_{t=l}^{r-1} W_t$ is the weight of this interval.

**Good Groups.** The idea of our proof is to keep $X_t$ within some bands, as depicted in Figure 4. In particular, we keep the curve above some bound $X_t \ge a$ such that $\text{loss}_t \le \epsilon = 2^{-a}$. There, we will keep the curve inside a band of width $3\beta$. At group boundaries, we even require the curve to be in the center-most $\beta$ of the band[1] so that we do not accumulate too much "debt" across groups. If necessary, we insert additional $\beta$ bits (= tasks) into the group, which moves the left end of the curve down by $\beta$ in expectation.

We aim for the drift of this random walk to be $\beta/2$ without inserting[1], so that we expect to insert every other group. We choose $\beta = \sqrt{b}\log b$ because the probability of the random walk leaving these bands in $\Theta(b)$ steps is low in $b$ because of the central limit theorem and properties of a normal distribution. If this is possible, we call a group *good*. In the rare cases when there are too many heavy keys or keys land clumped together, this insertion is not sufficient. We bump those groups' keys to the fallback data structure.

We do all these considerations assuming the maximal possible loss of $\epsilon$. In reality, when the loss is lower, the clamping of the $X_t$-curve at group boundaries to $X_{i,m_i} = \max\{X_{i+1,0}, a + \beta\}$ helps getting back to the center band. Clamping also brings the curve back to the bands when bumping, as the total absence of keys makes the curve go downwards.

**Parameters.** To make our analysis work, we make the following choices on the parameters introduced in Algorithm 1: For some large enough (determined later) $c$ and any $\delta \in (0, 2^{-cI_{\max}})$ we choose

[1] Ignoring additional terms depending on $\epsilon$ to compensate the loss for this high-level description.

- $b = 16/\delta$,
- $\beta = \sqrt{b}\log b$,
- $\epsilon = \delta/4$ and $a = \log_{½} \epsilon$,
- $\lambda = \frac{b+\beta/2}{H_0}(1-\epsilon)$.

An overview of all the parameters and variables can be found in Appendix A.

## 4.2 Probability of Good Groups

Now that we have given a sketch of our proof goal and introduced the required notation, we can formalize what it means for a group to be good. This is the most complicated part of our proof (see Figure 3). We order the proofs in increasing technicality so that more low-level proofs can be skipped easily.

Note that $X_t$ depends on $\ell_i$ (as well as $\ell_{i+1}, \ell_{i+2}, \dots$). Since we consider different values of $\ell_i$ in this section, we write $X_t^{(\ell)}$ for the value of $X_t$ in the case $\ell_i = \ell$ (but $\ell_{i+1}, \ell_{i+2}$ stay the same). We still write $X_t = X_t^{(\ell_i)}$.

▶ **Definition 3** (Good groups). *We define a group to be good if the event $\mathcal{G}_i := \mathcal{Q}_i \cap \mathcal{X}_i$ occurs. Let $a := \log_{½} \epsilon$. Its requirements are*

1. $\mathcal{Q}_i := \left\{\forall \ell \in \{0,1\}, j \in [m_i] : X_{i,j}^{(\ell)} \in [a, a+3\beta]\right\}$, *and*
2. $\mathcal{X}_i := \left\{\exists \ell \in \{0,1\} : X_{i,0}^{(\ell)} \in [a+\beta-\epsilon b, a+2\beta]\right\}$.

*We define a group to be bad $\mathcal{B}_i := \mathcal{G}_i^{\complement}$, if it is not good.*

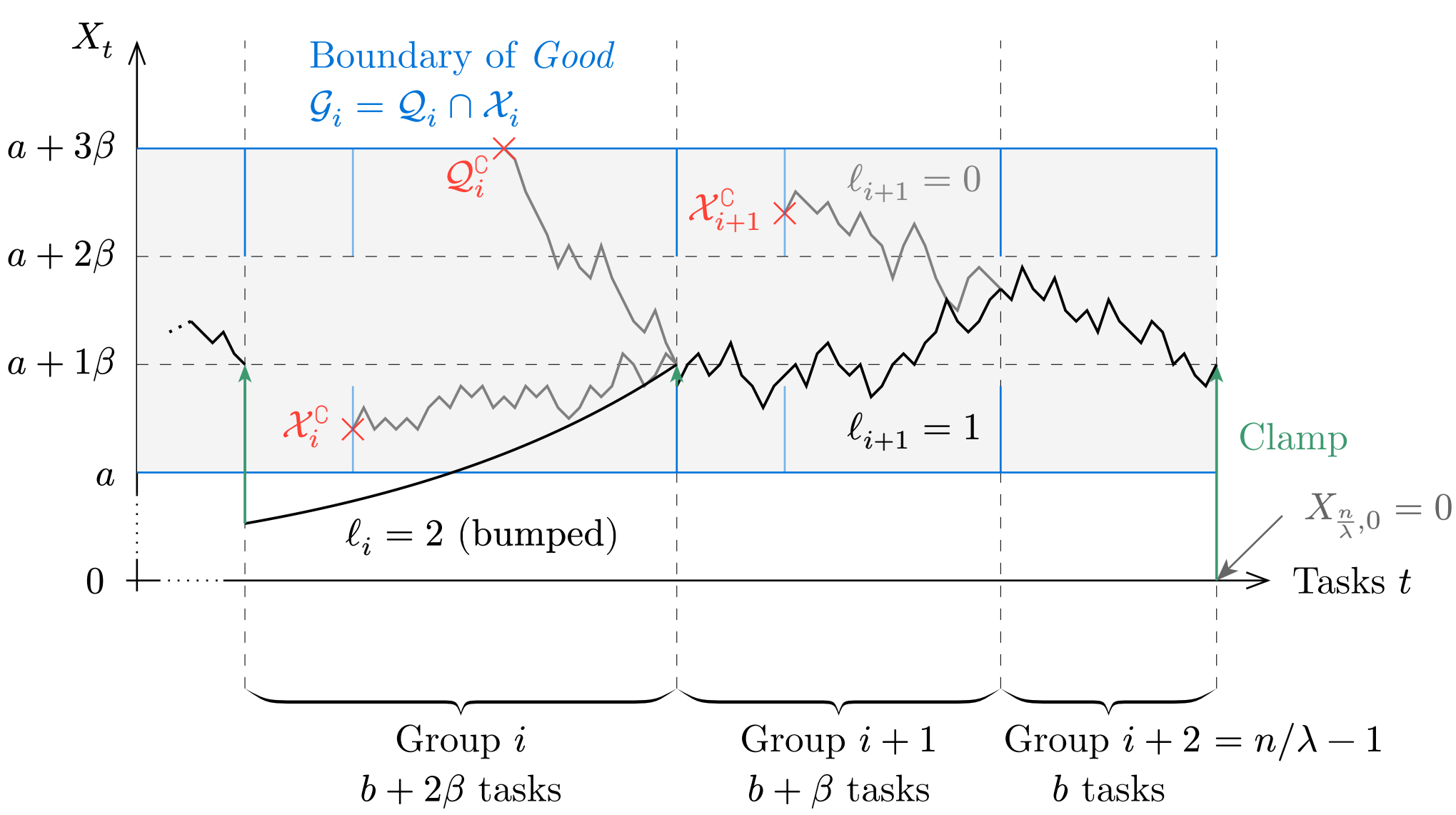

**Figure 4** This figure shows the $X_t$ curve over the tasks of three groups. In good groups, the curve must stay inside the three bands of width $\beta$, and must be inside the innermost band at group boundaries. The criterion for good groups $\mathcal{G}_i = \mathcal{Q}_i \cap \mathcal{X}_i$ is defined later on.

The curve starts at the end at $X_{n/\lambda,0} = 0$ and gets immediately clamped to the centermost band. The curve then follows a random walk based on the weight of the tasks $W_t$. If the curve is not inside the centermost band at a group boundary (e.g., in group $i+1$ the violation of $\mathcal{X}_{i+1}$) or the curve leaves the bands (e.g., violation of $\mathcal{Q}_i$ in group $i$) for $\ell_i = 0$, another curve is calculated for $\ell_i = 1$. Incrementing $\ell_i$ extends the length of the group by $\beta$. If this increased group size is still not sufficient (e.g., group $i$ where $\mathcal{X}_i$ is violated for $\ell_i = 1$), all the group's keys are bumped, and $\ell_i = 2$ is chosen to indicate this. As a consequence, the curve falls towards zero.

**Algorithm 3** Predicate function to check whether a group is *good* for a particular task assignment.

```
good(i, ℓ_i, (S_t)_{t≥(i,0)})
1  calculate X_{i,j}^{(ℓ_i)} from (S_t)_{t≥(i,0)} for j ∈ [b + ℓ_i β]
2  return ∀j ∈ [m_i] : X_{i,j}^{(ℓ_i)} ∈ [a, a + 3β]                      // In bands
      ∧ X_{i,0}^{(ℓ_i)} ∈ [a + β − εb, a + 2β]   // "Almost" in center band at group boundary
```

Figure 4 shows examples of how the requirements for these events may be violated. We define the *good* predicate used in Algorithm 1 to check this condition accordingly, see Algorithm 3. We can show that this enforces the invariant that $X_{i,m_i} \in [a+\beta, a+2\beta]$ over all groups, which allows us to consider each group individually.

▶ **Lemma 4** (Invariant). *For every group the invariant $X_{i,m_i} \in [a+\beta, a+2\beta]$ holds and $X_{0,0} \leq a+2\beta$.*

**Proof.** The clamping $\tilde{q}_{i,m_i} = \min\{q_{i+1,0}, \epsilon 2^{-\beta}\}$ ensures that $X_{i,m_i} = \log_{1/2} \tilde{q}_{i,m_i} \geq a+\beta$, since $a = \log_{1/2} \epsilon$. Therefore, we must only show $X_{i+1,0} \leq a + 2\beta$ for all $i \in \left[\frac{n}{\lambda}\right]$. We prove this using induction over the groups from right to left.

**Base $i = n/\lambda - 1$:** Since $q_{n/\lambda,0} = 1$ we have $X_{n/\lambda,0} = 0 \leq a + 2\beta$.

**Step $i+1 \rightsquigarrow i$:** Either group $i+1$ is a good group with $\ell_{i+1} \in \{0,1\}$ or a "bad" group where all the keys get bumped into the fallback and $\ell_{i+1} = 2$.

In the good case $\mathcal{X}_{i+1}$ holds and an $\ell_i \in \{0,1\}$ gets chosen by Algorithm 1 with the predicate of Algorithm 3 such that $X_{i+1,0} \leq a + 2\beta$.

In the "bad" case, all keys of the group get bumped into the fallback data structure. So the only increase is due to the loss, which is at most 1 bit per task. Thus $X_{i+1,0} \leq X_{i+1,m_{i+1}}$ and by induction hypothesis $X_{i+1,m_{i+1}} \leq a + 2\beta$.

The second part of the lemma follows by performing another induction step. ◀

We now show that bad groups are in fact rare.

▶ **Lemma 5** (Bad groups are rare). *The probability that group $i$ is bad is* $\Pr[\mathcal{B}_i] \leq b^{-\Omega(c)}$.

The key in the proofs for Lemma 5 is linear bounds on the change of $X_t$ during an interval of tasks $[l, r) = \{l, l+1, ..., r-1\}$ of width $\omega = r - l$:

$$\Delta_{lr} = X_l - X_r = W_{lr} - \omega + \sum_{t=l}^{r-1} \text{loss}_t$$

Here $W_{lr} = \sum_{t=l}^{r-1} W_t$ is the weight of all the keys inside said interval. As long as $X_t \geq a$, the loss is in $[0, \epsilon]$, and therefore $W_{lr} - \omega \leq \Delta_{lr} \leq W_{lr} - (1-\epsilon)\omega$. We can now apply the Chernoff bounds in Lemma 6 on $W_{lr}$ to that approximation, to show that a lot of change is unlikely.

▶ **Lemma 6.** *For some $\tau \in [0,1]$ let $W = \sum_{\alpha \in \Sigma} N^{(\alpha)} \log_{1/2} \rho_\alpha$ with $N^{(\alpha)} \sim \text{Bin}\left(\rho_\alpha n, \tau \frac{\lambda}{n}\right)$ independent. Let $I_{\max} \geq \log_{1/2} \rho_\alpha$ for all $\alpha \in \Sigma$ and $\mu := \mathbb{E}[W] = \tau\lambda H_0$.*

1. $\Pr\left[W \gtreqless x\right] \leq \exp\left[-\frac{1}{I_{\max}}\left(x \ln \frac{x}{\mu} - x + \mu\right)\right]$ $\;{(x>\mu) \atop (0<x<\mu)}$,
2. $\Pr[W \geq \mu + d] \leq \exp\left[-\frac{1}{I_{\max}} \frac{d^2}{2\mu + \frac{2}{3}d}\right]$ $\quad (d > 0)$,
3. $\Pr[W \leq \mu - d] \leq \exp\left[-\frac{1}{I_{\max}} \frac{d^2}{2\mu}\right]$ $\quad d \in (0, \mu)$.

Additionally, we need some inequalities on the variables, which follow from their definitions in Section 4.1.

▶ **Lemma 7** (Parameter Inequalities). *These inequalities hold:*

1. $16 \leq \beta \leq b$
2. $\epsilon < \frac{1}{4}$
3. $\beta < \lambda H_0 \leq \frac{3}{2}b$
4. $b\epsilon = 4$
5. $\epsilon\lambda H_0 \ / \ (1-\epsilon) \leq 3/8\beta$
6. $\frac{\log b}{I_{\max}} \geq c$

We provide the proof of Lemma 6 and Lemma 7 in Appendix B. With these helpful lemmata, we prove that bad groups are rare (Lemma 5) by reducing it to Lemma 8 and Lemma 9:

▶ **Lemma 8.** $\Pr[\mathcal{X}_i \mid \mathcal{Q}_i] \geq 1 - b^{-\Omega(c)}$.

▶ **Lemma 9.** $\Pr[\mathcal{Q}_i] \geq 1 - b^{-\Omega(c)}$.

**Proof of Lemma 5.** Using Lemma 8 and Lemma 9 we can show:

$$\Pr[\mathcal{B}_i] \leq \Pr[\mathcal{X}_i^{\complement} \cap \mathcal{Q}_i] + \Pr[\mathcal{Q}_i^{\complement}] \leq \Pr[\mathcal{X}_i^{\complement} \mid \mathcal{Q}_i] + \Pr[\mathcal{Q}_i^{\complement}] = b^{-\Omega(c)}.$$

◀

We can prove Lemma 8 immediately by considering all cases for $\mathcal{X}_i^{\complement}$ given $\mathcal{Q}_i$ and using a Chernoff bound on the group's weight:

**Proof of Lemma 8.** The weight of the keys in the group $W_i$ has expected value $\mu := \mathbb{E}[W_i] = \lambda H_0 = (1-\epsilon)(b + \beta/2)$.

In the case $\ell_i = \ell$, we denote the loss over the group as $L^{(\ell)}$. Thus,

$$X_{i,0}^{(\ell)} = X_{i,m_i} + W_i - (b + \ell\beta) + L^{(\ell)}.$$

Because of the condition $\mathcal{Q}_i$ the loss at each task is in $[0, \epsilon]$ and $L^{(\ell)} \in [0, \epsilon(b + \ell\beta)]$. Therefore, $X_{i,0}^{(1)} < X_{i,0}^{(0)}$, because

$$X_{i,0}^{(0)} - X_{i,0}^{(1)} = L^{(0)} - L^{(1)} + \beta > -\epsilon(b+\beta) + \beta = (1-\epsilon)\beta - \epsilon b \geq \frac{3}{4}16 - 4 > 0,$$

using Lemma 7, and $X_{i,0}^{(0)} - X_{i,0}^{(1)} = \beta + L^{(0)} - L^{(1)} \leq \beta + \epsilon b$ which is exactly the height of the admissible interval $I := [a + \beta - \epsilon b, a + 2\beta]$ of $\mathcal{X}_i$. So it is impossible that $X_{i,0}^{(1)}$ is above $X_{i,0}^{(0)}$, or that $X_{i,0}^{(1)}$ is below $I$ and $X_{i,0}^{(0)}$ is above. Therefore, the only two options for $\mathcal{X}_i^{\complement}$, i.e., $X_{i,0}^{(\ell)} \notin I$ for both $\ell = 0$ and $\ell = 1$, are $A := \left\{X_{i,0}^{(0)} < a + \beta - \epsilon b\right\}$ or $B := \left\{X_{i,0}^{(1)} > a + 2\beta\right\}$.

**Case A.** Because of $X_{i,m_i} \geq a + \beta$ by Lemma 4 and $L^{(0)} \geq 0$, the fact $X_{i,0}^{(0)} < a + \beta - \epsilon b$ implies

$$W_i = \underbrace{X_{i,0}^{(0)} - X_{i,m_i}}_{< -\epsilon b} + b - L^{(0)} \leq (1-\epsilon)b = \mu - (1-\epsilon)\frac{\beta}{2} < \mu - \frac{3}{8}\beta.$$

Since $\frac{3}{8}\beta < \beta < \lambda H_0 = \mu$ by Lemma 7, we can apply the Chernoff bound from Lemma 6, using $\Pr[\mathcal{Q}_i] \geq 1 - b^{-\Omega(c)} \geq 1/2$ (which holds for a sufficiently large $c$) at $\star$:

$$\Pr[A|\mathcal{Q}_i] \leq \frac{\Pr\left[W_i < \mu - \frac{3}{8}\beta\right]}{\Pr[\mathcal{Q}_i]} \overset{\star}{\leq} 2\exp\left(-\frac{1}{I_{\max}}\frac{\left(\frac{3}{8}\beta\right)^2}{2\mu}\right) \leq \exp\left(-\frac{1}{I_{\max}}\Omega\left(\frac{\beta^2}{b}\right)\right).$$

**Case B.** Because of $X_{i,m_i} \leq a + 2\beta$ by Lemma 4 and $L^{(1)} \leq \epsilon(b+\beta)$, the fact $X_{i,0}^{(1)} > a + 2\beta$ implies

$$W_i = \underbrace{X_{i,0}^{(1)} - X_{i,m_i}}_{>0} + (b+\beta) - L^{(1)} > (1-\epsilon)(b+\beta) = \mu + (1-\epsilon)\frac{\beta}{2} \geq \mu + \frac{3}{8}\beta$$

We can again apply a Chernoff bound from Lemma 6 using $\beta \leq b$:

$$\Pr[B|\mathcal{Q}_i] \le \frac{\Pr\left[X_{i,0}^{(1)} > a + 2\beta\right]}{\Pr[\mathcal{Q}_i]} \overset{\star}{\le} 2\Pr\left[W_i > \mu + \frac{3}{8}\beta\right] \le \exp\left(-\frac{1}{I_{\max}}\Omega\left(\frac{\beta^2}{b}\right)\right)$$

A union bound over the events $A$ and $B$ together with $\frac{\beta^2}{I_{\max} b} = \frac{\log^2 b}{I_{\max}} \ge c\log b$ by Lemma 7 yields the statement. ◀

For Lemma 9, which makes a statement about $X_t$ everywhere inside a group, we consider intervals of tasks. We show that, while still being inside the bands, $X_t$ is unlikely to grow or shrink by more than $\beta$ over some subinterval of tasks $[r, l)$ inside the group. This growth or shrinkage over some subinterval is a necessary condition for $\mathcal{Q}_i$ to be violated. We can further require the growing subinterval to have little loss. Let $\Delta_{lr} := X_l - X_r$ denote the delta over this subinterval of tasks.

▶ **Lemma 10.** *Let $\ell_i \in \{0, 1\}$. In an interval $[l, r)$ of tasks in group $i$*

$$\Pr[\Delta_{lr} > \beta \wedge \forall l \le t < r : q_t \le \epsilon] \le \exp\left(-\frac{1}{I_{\max}}\Omega\left(\frac{\beta^2}{b}\right)\right).$$

▶ **Lemma 11.** *Let $\ell_i \in \{0, 1\}$. In an interval $[l, r)$ of tasks in group $i$*

$$\Pr[\Delta_{lr} < -\beta] \le \exp\left(-\frac{1}{I_{\max}}\Omega\left(\frac{\beta^2}{b}\right)\right).$$

Using these lemmata in a union bound over all subintervals of tasks inside the group, we can then prove Lemma 9:

**Proof of Lemma 9.** Since $X_{i,m_i} \in [a + \beta, a + 2\beta]$ by Lemma 4, the event $\mathcal{Q}_i$, i.e., that $X_t \in [a, a + 3\beta]$ for all $t \in G_i$, must occur if there is no interval $[l, r)$ of tasks in the group $i$ where the change $\Delta_{lr}$ is less than $-\beta$ or more than $+\beta$. In fact, we only need to consider $\Delta_{lr} > \beta$ if $X_t > a$ for $l \le t < r$, because otherwise there would already be a different interval with $\Delta_{l'r'} < -\beta$.

By combining Lemma 10 and Lemma 11 with a union bound over all task intervals in the group, we get

$$\Pr[\mathcal{Q}_i^{\complement}] \le \binom{m_i}{2}\exp\left(-\frac{1}{I_{\max}}\Omega\left(\frac{\beta^2}{b}\right)\right) = \Theta(b^2)\exp\left(-\frac{1}{I_{\max}}\Omega(\log^2 b)\right) = b^{\Omega\left(-\frac{1}{I_{\max}}\log b + 2\right)}$$

This is $\le b^{-\Omega(c)}$ because by Lemma 7 $\frac{\log b}{I_{\max}} \ge c$, and thus the exponent is $\Omega(-c + 2) = -\Omega(c)$ for a large enough $c$. ◀

To prove the remaining lemmata, we transform the statements into a suitable form and apply a Chernoff bound.

**Proof of Lemma 10.** Let $W_{lr}$ be the weight of keys in the interval $[l, r)$ of tasks of width $\omega \le m_i = b + \ell_i\beta$ and $\mu := \mathbb{E}[W_{lr}] = \frac{\omega}{b+\beta\ell_i}\lambda H_0 = \omega(1-\epsilon)\frac{b+\beta/2}{b+\beta\ell_i}$.

Consider $d := \beta + (1 - \epsilon)\omega - \mu$:

$$d = \beta + (1-\epsilon)\omega\left(1 - \frac{b + \beta/2}{b + \beta\ell_i}\right) = \begin{cases} \beta - (1-\epsilon)\frac{\omega}{b}\frac{\beta}{2} \overset{\omega \le b}{\ge} \frac{\beta}{2} & \text{if } \ell_i = 0 \\ \beta + (1-\epsilon)\omega\left(\frac{\beta/2}{b+\beta}\right) \ge \frac{\beta}{2} & \text{if } \ell_i = 1. \end{cases}$$

Using this at $*$ and the fact that $\mu \le \lambda H_0 \le \frac{3}{2}b$ from Lemma 7 at $**$ we can prove this lemma's statement:

$$\Pr[\Delta_{lr} > \beta \wedge \forall l \le t < r : q_t \le \epsilon] \le \Pr[W_{lr} - \omega(1-\epsilon) > \beta] = \Pr[W_{lr} > \mu + d]$$
$$\overset{*}{\le} \Pr\left[W_{lr} > \mu + \frac{\beta}{2}\right] \overset{\text{Lemma 6}}{\le} \exp\left(-\frac{1}{I_{\max}} \frac{(\beta/2)^2}{2\mu + \frac{2}{3}\frac{\beta}{2}}\right) \overset{**}{\le} \exp\left(-\frac{1}{I_{\max}}\Omega\left(\frac{\beta^2}{b}\right)\right).$$

◀

**Proof of Lemma 11.** We consider an interval of tasks $[l, r)$ in group $i$ of width $\omega = r - l \le m_i = b + \ell_i \beta$. We only need to consider the case of $\omega > \beta$ since otherwise the event $\{W_{lr} < \omega - \beta\} = \varnothing$ because $W_{lr} \ge 0$. We will use $d := \mu - \omega + \beta$ for Lemma 6.3. We claim that $\frac{1}{8}\beta < d < \mu$. Because of $\omega > \beta$ is $d = \mu - (\omega - \beta) < \mu$. If we use the fact that $\mu \le \lambda H_0$ and $\frac{\epsilon}{1-\epsilon}\lambda H_0 \le \frac{3}{8}\beta$ by Lemma 7 at $*$ we can show

$$d = \beta + \mu - \omega = \beta + \omega\left(\frac{b + \beta/2}{b + \ell_i\beta} - 1\right) - \frac{\epsilon}{1-\epsilon}\mu \overset{*\ \&\ \ell_i \le 1}{\ge} \beta + \omega\left(\frac{b + \beta - \beta/2}{b + \beta} - 1\right) - \frac{3}{8}\beta$$
$$\overset{\omega \le b + \beta}{\ge} \beta - \frac{\beta}{2} - \frac{3}{8}\beta = \frac{1}{8}\beta.$$

Now we use the Chernoff bound from Lemma 6.3 together with the facts that $\mu \le \lambda H_0 \le \frac{3}{2}b$, and $d > \frac{1}{8}\beta$:

$$\Pr[\Delta_{lr} < -\beta] \le \Pr[W_{lr} - \omega < -\beta] = \Pr[W_{lr} \le \mu - d]$$
$$\le \exp\left(-\frac{1}{I_{\max}}\frac{d^2}{2\mu}\right) \le \exp\left(-\frac{1}{I_{\max}}\frac{\left(\frac{1}{8}\beta\right)^2}{3b}\right) \le \exp\left(-\frac{1}{I_{\max}}\Omega\left(\frac{\beta^2}{b}\right)\right).$$

◀

## 4.3 Bad Group Handling

Now that we know that bad groups are rare, we show some properties of the fallback data structure for the proof of Theorem 2. We handle bad groups by bumping them into a fallback data structure $F$. We know from Lemma 5 that only a fraction of $b^{-\Omega(c)}$ of the groups is bad, so we have some freedom in how to handle those. For simplicity of the analysis, we use a simple minimal perfect hashing–based compressed retrieval data structure like the one described as a side note in [1]. Note that during our experiments we use a different fallback mechanism that is simpler to implement and we expect to be as space efficient but harder to analyze. In the end, we prove the following theorem:

▶ **Theorem 12.** *The fallback data structure has linear construction time in $|S_F|$, queries take constant time, and its expected size is*

$$\mathbb{E}[|F|] = \mathcal{O}\left(nb^{-\Omega(c)}\left(H_0 + \frac{H_0}{I_{\min}}\log I_{\max}\right) + |\Sigma|\right).$$

We split its proof into two parts. First, we show properties of the fallback dependent on the number and weight of bumped keys. Then we show how many keys will actually get bumped in expectation, and what their weight is.

▶ **Corollary 13** (Fallback like Lemma 2.1. in [1]). *Let $N_F := |S_F|$ be the number of keys bumped into the fallback and $W_F := \sum_{x \in S_F} \log_{1/2} \rho_{f(x)}$ their weight according to the original value distribution. In expectation, the fallback data structure can be constructed in $\mathcal{O}(N_F)$, has constant query time, and a space usage of $W_F + N_F \log\left(\frac{W_F}{N_F} + 1\right) + \mathcal{O}(N_F + |\Sigma|)$ bits.*

**Proof.** We use a data structure based on an MPHF and a Huffman code from Lemma 2.1 in [1]. It takes up space $N_F H_F + N_F \log(H_F + 1) + \mathcal{O}(N_F + |\Sigma|)$ with $H_F$ being the Shannon entropy of the fallback values $(f(x))_{x\in S_F}$. This entropy may be different than the entropy $H_0$, as keys with different weight might be bumped with a different probability. Still we can upper bound this space usage with $W_F + N_F \log\left(\frac{W_F}{N_F}+1\right) + \mathcal{O}(N_F + |\Sigma|)$, because $N_F H_F \le W_F$ since cross-entropy is at least entropy by Gibbs' inequality [4, 9].

If we choose an MPHF with constant query time and expected linear construction like [17], which only needs $\mathcal{O}(N)$ bits of space for $N$ keys, we get constant query time and linear construction time for the whole fallback data structure, since the Huffman codes and the prefix sum data structure already have linear construction and constant query time. ◀

▶ **Lemma 14.** *The expected number of bumped keys $N_F := |S_F|$ and their weight $W_F := \sum_{x\in S_F} \log_{1/2} \rho_{f(x)}$ are*

$$\mathbb{E}[N_F] \le n\frac{H_0}{I_{\min}} b^{-\Omega(c)}, \text{ and } \mathbb{E}[W_F] \le nH_0 b^{-\Omega(c)}.$$

**Proof of Lemma 14.** First, we look at a single bad group $i$, i.e., one with the bad event $\mathcal{B}_i$. $W_i > b + 2\beta$ is sufficient for a group to be bad because $X_{i,0} - X_{i,m_i} \ge W_i - (b + \ell_i\beta)$:

$$\mathcal{G}_i \implies \mathcal{X}_i \implies W_i \le X_{i,0} - X_{i,m_i} + b + \ell_i\beta \le a + 2\beta - (a+\beta) + b + \beta = b + 2\beta.$$

Thus, using $T := b + 2\beta \le \mathcal{O}(b)$ and $\{W_i > T\} \subseteq \mathcal{B}_i$ we obtain the following chain of inequalities:

$$\mathbb{E}\left[W_i \mathbb{1}_{\mathcal{B}_i}\right] = \mathbb{E}\left[W_i \mathbb{1}_{\mathcal{B}_i \cap \{W_i \le T\}}\right] + \mathbb{E}\left[W_i \mathbb{1}_{\mathcal{B}_i \cap \{W_i > T\}}\right] \le T \Pr[\mathcal{B}_i] + \mathbb{E}\left[W_i \mathbb{1}_{\{W_i > T\}}\right]$$

$$\overset{\text{Cauchy-Schwarz}}{\le} T\Pr[\mathcal{B}_i] + \sqrt{\mathbb{E}[W_i^2]\Pr[W_i > T]} \overset{*}{\le} \mathcal{O}(b)b^{-\Omega(c)} + \mathcal{O}(b)b^{-\Omega(c/2)} = \mathcal{O}(b)b^{-\Omega(c)}.$$

Inequality $*$ uses $\Pr[W_i > T] \le \Pr[\mathcal{B}_i] \le b^{-\Omega(c)}$ by Lemma 5 and

$$\mathbb{E}[W_i^2] = \operatorname{Var}(W_i) + \mathbb{E}[W_i]^2 \overset{\dagger}{=} \mathcal{O}(b\log b) + \mathcal{O}(b^2) = \mathcal{O}(b^2).$$

For the bound on the variance in † we use that $W_i = \sum_{\alpha\in\Sigma} \log_{1/2} \rho_\alpha N_i^{(\alpha)}$ with $N_i^{(\alpha)} \overset{\text{ind.}}{\sim} \operatorname{Bin}\left(\rho_\alpha n, \frac{\lambda}{n}\right)$ for $\alpha \in \Sigma$, and $\log_{1/2} \rho_\alpha \le I_{\max} = \mathcal{O}(\log b)$ by Lemma 7.6. Therefore,

$$\operatorname{Var}(W_i) \overset{\text{ind.}}{=} \sum_{\alpha\in\Sigma} (\log_{1/2} \rho_\alpha)^2 \rho_\alpha n \frac{\lambda}{n}\left(1 - \frac{\lambda}{n}\right) \le I_{\max}\lambda H_0 \le \mathcal{O}(b\log b).$$

Note that since $I_{\min}$ is the minimum weight of a key, we have $N_i \le W_i / I_{\min}$ and we get $\mathbb{E}\left[N_i \mathbb{1}_{\mathcal{B}_i}\right] \le \mathcal{O}(b/I_{\min}) b^{-\Omega(c)}$ by linearity.

Summing up over all groups and using $\lambda = \Theta(b/H_0)$, we get

$$\mathbb{E}[W_F] = \sum_{i\in[\frac{n}{\lambda}]} \mathbb{E}\left[W_i \mathbb{1}_{\mathcal{B}_i}\right] \le \frac{n}{\lambda}\cdot\mathcal{O}(b)b^{-\Omega(c)} = nH_0 b^{-\Omega(c)}, \text{ and}$$

$$\mathbb{E}[N_F] = \sum_{i\in[\frac{n}{\lambda}]} \mathbb{E}\left[N_i \mathbb{1}_{\mathcal{B}_i}\right] \le \frac{n}{\lambda}\cdot\mathcal{O}(b/I_{\min})b^{-\Omega(c)} = n\frac{H_0}{I_{\min}} b^{-\Omega(c)}.$$

◀

These results now allow us to prove the theorem about the fallback.

**Proof of Theorem 12.** We plug Lemma 14 into Corollary 13 to get

$$\mathbb{E}[|F|] = \mathbb{E}\left[W_F + N_F \log\left(\frac{W_F}{N_F} + 1\right) + \mathcal{O}(N_F + |\Sigma|)\right]$$

$$\overset{*}{\le} \mathbb{E}[W_F] + \mathbb{E}[N_F(\log(I_{\max}+1))] + |\Sigma| \overset{\text{Lemma 14}}{\le} \mathcal{O}\left(nb^{-\Omega(c)}\left(H_0 + \frac{H_0}{I_{\min}}\log I_{\max}\right) + |\Sigma|\right).$$

The inequality $*$ uses $I_{\max} \ge \frac{W_F}{N_F}$. ◀

## 4.4 Summing up

We now want to prove the main theorem about the construction-time–space tradeoff and query time. We first determine the number of total consensus tasks:

▶ **Lemma 15** (Length of Consensus vector). $\mathbb{E}\left[\sum_{i\in[n/\lambda]} b + \ell_i\beta\right] \leq nH_0(1+\mathcal{O}(\delta))$.

**Proof of Lemma 15.** We separate the sum $M := \sum_{i\in[n/\lambda]} b + \ell_i\beta$ into good and bad groups. In good groups $X_{i,0} - X_{i,m_i} \leq W_i - (b+\ell_i\beta)(1-\epsilon)$ holds, because of the bounded loss. This inequality allows the following bound:

$$M_G := \sum_{i\in[n/\lambda]\wedge\mathcal{G}_i} b + \ell_i\beta \leq \frac{1}{1-\epsilon}\sum_{i\in[n/\lambda]\wedge\mathcal{G}_i} W_i - X_{i,0} + X_{i,m_i}$$

In good groups clamping moves $X_t$ at most by $\epsilon b$ because of $\mathcal{X}_i$: $X_{i,0} \geq a + \beta - b\epsilon$ and thus $X_{i-1,m_{i-1}} = \max\{a+\beta, X_{i,0}\} \leq \max\{X_{i,0} + b\epsilon, X_{i,0}\} \leq X_{i,0} + \epsilon b$. Further, we can add bad groups back in, as for bad groups $X_{i,m_i} \geq a+\beta$ (Lemma 4) and $X_{i,0} \leq X_{i,m_i}$ as all keys are bumped and loss is at most one. Thus, $X_{i-1,m_i} = \max\{a+\beta, X_{i,0}\} \leq \max\left\{X_{i,m_i}, X_{i,m_i}\right\} = X_{i,m_i}$. For convenience we introduce $X_{-1,m_{-1}} := \max\{a+\beta, X_{0,0}\}$.

$$\begin{aligned}(1-\epsilon)M_G &\leq \sum_{i\in[n/\lambda]\wedge\mathcal{G}_i} W_i - X_{i-1,m_{i-1}} + X_{i,m_i} + b\epsilon \leq \sum_{i\in[n/\lambda]} W_i - X_{i-1,m_{i-1}} + X_{i,m_i} + b\epsilon \\ &\overset{🔭}{\leq} nH_0 - X_{-1,m_{-1}} + X_{n/\lambda-1,m_{n/\lambda-1}} + \frac{n}{\lambda}b\epsilon\end{aligned}$$

The inequality 🔭 folds the telescope sum, and $\sum_{i\in[n/\lambda]} W_i = nH_0$ since that is just the weight of all keys in $S$. We know $X_{-1,m_{-1}} \geq a+\beta$ by Lemma 4 (extending it one step further) and

$$X_{n/\lambda-1,m_{n/\lambda-1}} = \log_{½}\tilde{q}_{n/\lambda-1,m_{n/\lambda-1}} = \log_{½}\min\left\{\epsilon 2^{-\beta}, q_{n/\lambda,0}\right\} = \log_{½}\min\left\{\epsilon 2^{-\beta}, 1\right\} = a+\beta,$$

and therefore $X_{n/\lambda-1,m_{n/\lambda-1}} - X_{-1,m_{-1}} \leq 0$. Thus in total with $\lambda = \Theta(b/H_0)$ we get

$$(1-\epsilon)M_G \leq nH_0 + \frac{n}{\lambda}b\epsilon = nH_0(1+\Theta(\epsilon)).$$

For bad groups, we know $\ell_i = 2$ and thus $M_B := \sum_{i\in[n/\lambda]} \mathbb{1}_{\mathcal{B}_i}\cdot(b+2\beta)$ which yields in expectation

$$\mathbb{E}[M_B] = \sum_{i\in[n/\lambda]} \Pr[\mathcal{B}_i](b+2\beta) \overset{\text{Lemma 5}}{\leq} \mathcal{O}\left(\frac{n}{\lambda}b^{-\Omega(c)}b\right) = \mathcal{O}\left(nH_0 b^{-\Omega(c)}\right).$$

Thus, in total we get with $(1+\Theta(\epsilon))(1-\epsilon)^{-1} = 1+\Theta(\epsilon)$ and $b^{-\Omega(c)} = O\left(\frac{1}{b}\right) = O(\delta)$ for a large enough $c$

$$\mathbb{E}[M] = \mathbb{E}[M_G + M_B] \leq \left(1+\Theta(\epsilon) + b^{-\Omega(c)}\right)\cdot nH_0 \leq nH_0(1+\mathcal{O}(\delta)).$$

◀

Using this, we prove the main theorem:

**Proof of Theorem 2 (Main Theorem).** **Query time** is constant as select queries are constant time, hashing can be done in constant time in our model, and accessing the alias table takes constant time (see also Section 3).

**Space usage** consists of four parts. The insertion vector $I$ length is $|I| \leq \mathcal{O}\left(\frac{n}{\lambda}\right) \leq \mathcal{O}(\delta n H_0)$, as for every group we have one separator bit and at most $\ell_i \leq 2$ inserted bits. Overhead for the select data structure is also linear. The fallback data structure uses $\mathbb{E}[|F|] := \mathcal{O}\left(nb^{-\Omega(c)}\left(H_0 + \frac{H_0}{I_{\min}}\log I_{\max}\right) + |\Sigma|\right)$ as per Theorem 12. The consensus vector has

size $\mathbb{E}\left[\sum_{i\in[n/\lambda]} m_i\right] \leq nH_0(1+\mathcal{O}(\delta))$ as per Lemma 15. Additionally, Consensus needs a root seed $R \sim \text{Geo}_0(q_{0,0})$ to encode retries when backtracking beyond task 0. For its space usage, we get as in [17]

$$\mathbb{E}[\log(1+R)] \overset{*}{\leq} \log \mathbb{E}[1+R] = \log \frac{1}{q_{0,0}} \leq X_{0,0} \overset{**}{\leq} a + 2\beta = \mathcal{O}\left(\sqrt{1/\delta}\log 1/\delta\right).$$

Here, $*$ uses Jensen's inequality [15] and $**$ follows from Lemma 4. Clamping increases the curve, and thus we get an upper bound. The alias table needs $\mathcal{O}(|\Sigma|)$ space (see Section 3). In total with $\epsilon = \Theta(\delta)$ and a large enough choice of $c$ such that $b^{-\Omega(c)} \leq \mathcal{O}\left(\frac{1}{b}\right) = \mathcal{O}(\delta)$ we get the claimed space usage of

$$\mathbb{E}[\text{Space}] \leq nH_0 + \delta n \cdot \mathcal{O}(H_0 + H_0/I_{\min} \cdot \log I_{\max}) + \mathcal{O}\left(\sqrt{1/\delta}\log 1/\delta + |\Sigma|\right).$$

**Construction time.** Observing Algorithm 1, the first phase until Line 10 takes linear time in the number of tasks in the end, which are $\mathcal{O}(nH_0)$ (Lemma 15). In particular, evaluating the partitioning and hash functions takes constant time in our hashing model. Further, checking whether a group is good as specified in Algorithm 3 takes time $\mathcal{O}(b + \ell_i\beta)$ when $X_{i+1,0}$ is remembered.

The alias table can be constructed in $\mathcal{O}(|\Sigma|) \leq \mathcal{O}(n)$ (see Section 3).

The construction of consensus is dominated by the number of hash function evaluations (apart from iterating over all tasks), each of which takes constant time. As tasks may have different amounts of keys mapped to them, it is easier to sum up over the keys. From the definition of $q_{t+1}$ we know a key $k$ in task $t$ is evaluated $T_k \sim \text{Geo}_1(p_t q_{t+1})$ times, see Section 2.1. Using our upper bound on $X_t \leq a + 3\beta$ in good groups and that bad groups contain no keys, we get

$$\epsilon 2^{-3\beta} \leq q_t = 1 - (1 - p_t q_{t+1})^2 \leq 2p_t q_{t+1}$$

and thus $\mathbb{E}[T_k] \leq \mathcal{O}(8^\beta/\epsilon)$. Summing over all $n$ independently successful keys, we get $\mathbb{E}[\#\text{hash evaluations}] \leq \mathcal{O}\left(\frac{n}{\epsilon}8^\beta\right) \leq n\exp\left(\Theta\left(\sqrt{1/\delta}\log 1/\delta\right)\right)$.

The fallback construction takes time linear in the number of bumped keys (Theorem 12), which is $\mathbb{E}[N_F] = \mathcal{O}\left(nH_0/I_{\min} \cdot b^{-\Omega(c)}\right)$ (Lemma 14). The select data structure can be constructed in $|I| \leq \mathcal{O}(\delta n H_0)$ time (see Section 3). Summing this all up with a large enough choice of $c$ such that $b^{-\Omega(c)} = \mathcal{O}(1/b) = \mathcal{O}(\delta)$ yields the stated construction time of

$$\mathbb{E}[\text{Construction Time}] \leq n \cdot \left(2^{\mathcal{O}\left(\sqrt{1/\delta}\log 1/\delta\right)} + \mathcal{O}(H_0 + H_0/I_{\min})\right).$$

◀

# 5 Experiments

In addition to the theoretical algorithm, which we analyzed thoroughly, we also present an implementation of a slightly modified algorithm. In this section, we explain the differences of this modified version and briefly present the experimental results of comparing it with two main competitors.

## 5.1 Difference to the Theoretical Algorithm

For our implementation, we make some modifications to the algorithm as described in Algorithm 1 and Section 4. The main difference is that we do not use a fallback data structure and instead insert $\ell_i \in \mathbb{N}_0$ times $\beta$ additional tasks until the group is considered good. Since bad groups are rare, this does not make a huge difference for the performance, but simplifies the implementation—just like the fallback simplified the analysis. Additionally, we use Select9 [25, 26] and the AES-based ahash [16] as a practical implementation of the select data structure and the hash functions together with some alias table optimizations [13]. We first hash all keys to 64-bit integer hash codes, so that we do not need to evaluate hash functions on large keys, e.g., long strings.

**Table 1** Practical choices of the parameters we use for the experiment. The maximum allowed difficulty $\log_{½} q_t$ of a task $t$ inside a group is $D$, and $D_b$ is the maximum allowed at a group border. The parameter $b$ is the initial size of a group, $\beta$ the insertion increment, and $\lambda H_0$ is the average weight inside groups.

| | $D$ | $D_b$ | $b$ | $\beta$ | $\lambda H_0$ |
|---|---|---|---|---|---|
| **A** | 8 | 6 | 30 | 4 | 35 |
| **B** | 12 | 11 | 30 | 4 | 35 |
| **C** | 16 | 12 | 30 | 8 | 35 |

We do not use any clamping and just look at the $\log_{½} q_t$. Additionally, we no longer consider a group bad if the $\log_{½} q_t$ is too low. This was important for bounding the loss in the analysis, but in practice there are just some easy tasks which do not cause problems. Instead, a group is considered bad if for any task $t$ the difficulty $\log_{½} q_t$ exceeds a threshold $D$ or the task at the beginning of a group has difficulty $\log_{½} q_{i,0} > D_b$. In the analysis, the asymptotic choices for these thresholds were $D = a + 3\beta$ and $D_b = a + 2\beta$.

In addition to the difficulty thresholds $D$ and $D_b$, our algorithm is parametrized by the initial group size $b$, the insertion increment $\beta$, and the expected group load $\lambda H_0$, which already appear in the analysis. As our analysis only makes asymptotic statements but constant factors are relevant in practice, we independently optimized the choice of these parameters for our experiments. We chose three different maximum intra-group task difficulties $D \in \{8, 12, 16\}$ which lead to still practical construction time and used manual grid search to identify the values for the other parameters which minimize the space. These choices are shown in Table 1.

## 5.2 Setup

We run our experiment on a Intel(R) Xeon(R) Gold 6314U CPU @ 2.30GHz.

**Compiler.** Our implementation is written in Rust, compiled with rustc 1.97.1 (8bab26f4f 2026-07-14) in release mode with `lto=true`, `codegen-units=1` and `target-cpu=native`. Our implementation can be found on GitHub [21] as well as our benchmark setup [22].

**Benchmark.** For our benchmarks we use synthetic data sets with $n = 10{,}000{,}000$ keys each, and values randomly generated according to three different types of distributions: $\mathcal{U}([k])$ is uniformly drawn from $k \in \{3, 23\}$ values, $\mathrm{Ber}(p)$ is a $\{0, 1\}$–distribution with the probability of 1 being $p \in \{0.1, 0.25, 0.5\}$, and a truncated geometric distribution[2] $\mathrm{Geo}(0.2) \bmod 10$. We choose these distributions as they contain different aspects that may be difficult for a CSF: Uniform distributions provide a baseline and also show behavior if there are only heavy keys, unbalanced Bernoulli distributions are not well encodable with a Huffman code, and the truncated geometric distribution covers a wide but bounded range of key weights.

---

[2] A value $Y \sim \mathrm{Geo}(0.2) \bmod 10$ is in $\{0, 1, \ldots, 9\}$ and $\Pr[Y = k] = \Pr[Z \equiv k \pmod{10}] = \Pr[Z = k \mid Z < 10]$ for $Z \sim \mathrm{Geo}(0.2)$, so $Y$ is distributed almost geometrically while avoiding the uncommon large values.

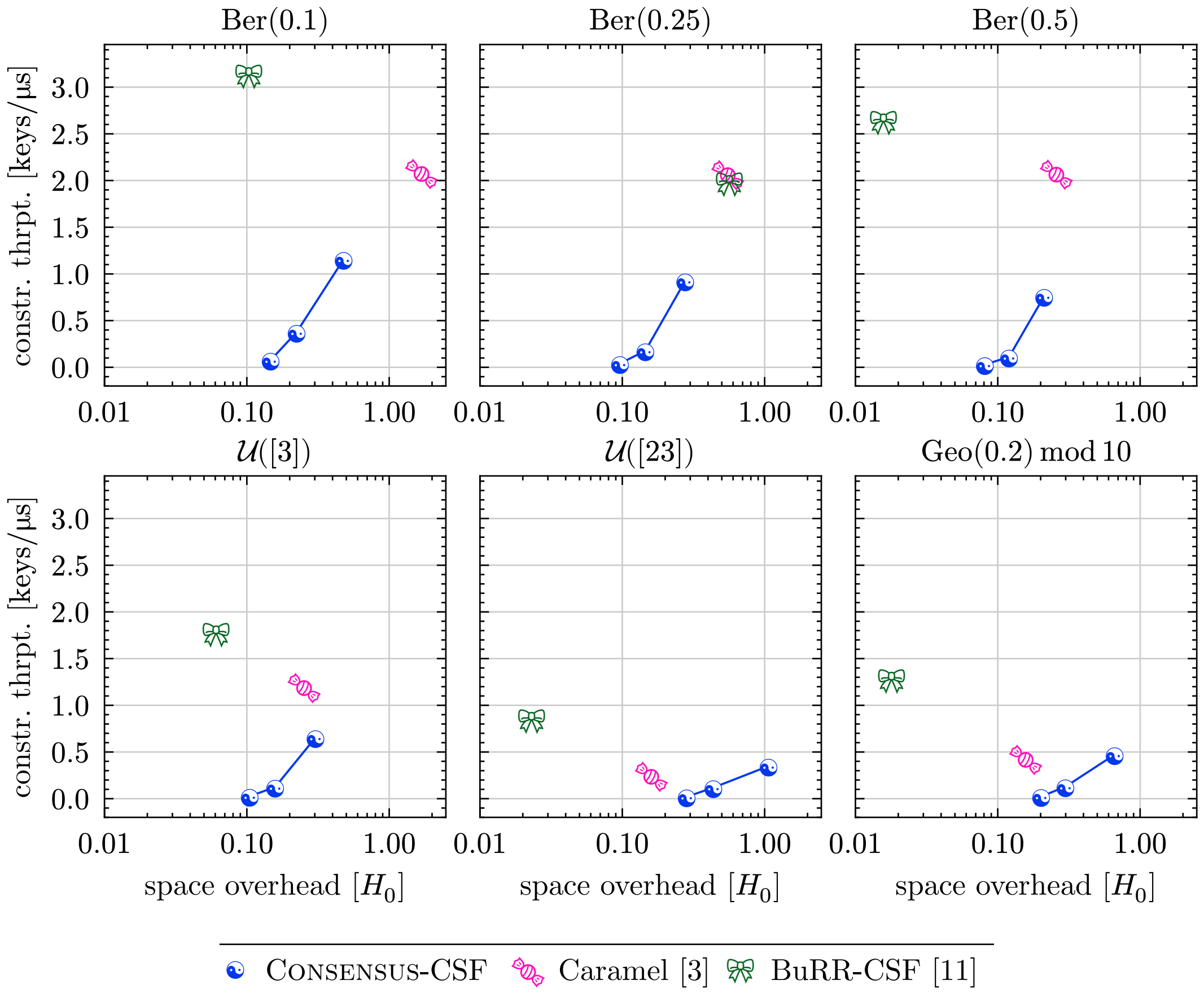

**Figure 5** Tradeoff plots for the different distributions, $n = 10{,}000{,}000$. The y-axis is the construction throughput in keys per µs, and the x-axis is the space overhead relative to $H_0$ with a logarithmic scale. CONSENSUS curve shows configurations A, B and C from Table 1, where configuration A has the highest overhead. Concrete values can be found in Table 2. Top left is better.

**Competitors.** We compare our implementation against Caramel [3] as the most advanced and easiest-to-integrate implementation of the *linear-equation* family of algorithms [12]. Further, we compare against BuRR-CSF [11]. We run all algorithms without parallelism.

## 5.3 Results

Figure 5 shows the different tradeoffs between construction time and space overhead. Concrete values are provided in Table 2.

**Configurations.** The configurations of CONSENSUS-CSF (Table 1) with larger $D$, i.e., those accepting more difficult tasks, have lower space overhead but take longer to construct. For example on Ber(0.25) the space overheads are 28%, 15%, and 10% with construction time per key of 1 µs, 6 µs, and 25 µs.

**Construction.** The construction of CONSENSUS-CSF is significantly slower than the competitors. For example, in the Ber(0.1) setting, construction time is 0.9 µs, 2.7 µs, 13.0 µs per key for CONSENSUS-CSF (A, B, C). Caramel takes 0.5 µs, BuRR-CSF takes 0.3 µs per key.

**Table 2** Benchmarking results for $n = 10{,}000{,}000$. Space overhead is relative to $H_0$. All values are per key. Gray ± values are standard deviations of the query time with 1000 samples for different keys.

| | $\mathcal{U}([3])$ | $\mathcal{U}([23])$ | Ber(0.1) | Ber(0.25) | Ber(0.5) | Geo(0.2) mod 10 |
|---|---|---|---|---|---|---|
| $H_0$ | 1.58 | 4.52 | 0.47 | 0.81 | 1.00 | 3.06 |
| **Consensus-CSF A** $D = 8$ | | | | | | |
| Construction [µs] | 1.5 | 2.9 | 0.9 | 1.1 | 1.3 | 2.1 |
| Query time [ns] | 67.7 ±12.6 | 96.3 ±22.6 | 52.6 ±10.7 | 58.8 ±11.5 | 58.9 ±11.7 | 81.8 ±16.6 |
| Space usage [bit] | 2.07 | 9.34 | 0.69 | 1.04 | 1.21 | 5.08 |
| Space overhead [%] | 30.3 | 106.5 | 47.8 | 27.7 | 21.2 | 66.2 |
| Space of $I$ [bit] | 0.22 | 1.53 | 0.09 | 0.11 | 0.12 | 0.71 |
| Space of $C$ [bit] | 1.84 | 7.81 | 0.60 | 0.93 | 1.09 | 4.38 |
| **Consensus-CSF B** $D = 12$ | | | | | | |
| Construction [µs] | 8.1 | 8.4 | 2.7 | 5.6 | 9.0 | 7.8 |
| Query time [ns] | 64.5 ±12.4 | 82.8 ±18.2 | 50.8 ±9.0 | 54.8 ±10.8 | 56.8 ±11.7 | 73.1 ±13.4 |
| Space usage [bit] | 1.84 | 6.50 | 0.57 | 0.93 | 1.12 | 3.97 |
| Space overhead [%] | 15.8 | 43.6 | 22.3 | 14.5 | 12.0 | 29.9 |
| Space of $I$ [bit] | 0.17 | 0.80 | 0.06 | 0.08 | 0.10 | 0.42 |
| Space of $C$ [bit] | 1.67 | 5.69 | 0.52 | 0.85 | 1.02 | 3.55 |
| **Consensus-CSF C** $D = 16$ | | | | | | |
| Construction [µs] | 37.7 | 48.9 | 13.0 | 24.7 | 33.9 | 42.0 |
| Query time [ns] | 61.3 ±11.6 | 76.1 ±13.9 | 50.3 ±8.6 | 52.3 ±9.6 | 62.0 ±9.7 | 65.0 ±13.6 |
| Space usage [bit] | 1.75 | 5.80 | 0.54 | 0.89 | 1.08 | 3.68 |
| Space overhead [%] | 10.5 | 28.3 | 14.7 | 9.6 | 8.1 | 20.2 |
| Space of $I$ [bit] | 0.11 | 0.43 | 0.03 | 0.05 | 0.06 | 0.26 |
| Space of $C$ [bit] | 1.64 | 5.37 | 0.50 | 0.84 | 1.02 | 3.42 |
| **Caramel** [3] | | | | | | |
| Construction [µs] | 0.8 | 3.7 | 0.5 | 0.5 | 0.5 | 2.2 |
| Query time [ns] | 60.4 ±14.3 | 99.9 ±22.5 | 43.6 ±9.4 | 43.8 ±9.5 | 43.8 ±8.6 | 80.7 ±19.0 |
| Space usage [bit] | 1.98 | 5.25 | 1.26 | 1.26 | 1.26 | 3.54 |
| Space overhead [%] | 25.2 | 16.0 | 168.6 | 55.1 | 25.8 | 15.7 |
| **BuRR-CSF** [11] | | | | | | |
| Construction [µs] | 0.6 | 1.2 | 0.3 | 0.5 | 0.4 | 0.8 |
| Query time [ns] | 71.5 ±13.7 | 102.9 ±20.6 | 73.8 ±13.1 | 67.1 ±12.3 | 48.9 ±10.8 | 94.5 ±17.8 |
| Space usage [bit] | 1.68 | 4.63 | 0.52 | 1.27 | 1.02 | 3.11 |
| Space overhead [%] | 6.1 | 2.3 | 10.3 | 56.5 | 1.6 | 1.8 |

**Space.** BuRR-CSF is the most space-efficient algorithm with an overhead of, e.g., just 10.3% over the empirical entropy for Ber(0.1). Only with the Ber(0.25) values is its overhead at 56.5% worse than all other algorithms. We suspect this is a mistake in the implementation.

With its large difficulty bound $D$, configuration $C$ reaches an overhead of 14.7%, close to BuRR-CSF in the Ber(0.1) case. The other configurations have worse overheads (47.8%, 22.3%). Caramel is particularly bad in this setting (168.6%). Configurations A and B undercut Caramel with regard to space usage on inputs with few values $(\mathrm{Ber}(p), \mathcal{U}([3]))$ but fall behind Caramel when there are more and in particular heavy values $(\mathcal{U}([23]), \mathrm{Geo}(0.2) \bmod 10)$. This may be improved by splitting heavy keys into multiple parts; see future work (Section 6.1).

As expected, the Consensus vector $C$ part of the space usage is always already larger than $H_0$. Space used for the insertion vector $I$ is always "wasted" as it does not encode any zeroth-order information.

**Query.** The query times are similar. The query times in the Ber(0.1) case of Consensus-CSF are 52.6 ns, 50.8 ns, and 50.3 ns. Caramel is slightly faster (43.6 ns), and BuRR-CSF is slower (73.8 ns). Query times have a positive correlation with the number of values $|\Sigma|$ in all implementations.

**Theory–Practice Gap.** Although our algorithm beats the above-shown competitors regarding space usage in theory—being able to achieve an arbitrarily low space overhead —in practice, having a construction time cost of $n \exp\left(\tilde{\mathcal{O}}\left(\sqrt{1/\delta}\right)\right)$ for $nH_0(1+\delta)$ space usage is still too expensive for reaching small space overheads. Our asymptotic analysis only shows its effects for small $\delta$ where construction time already is impractical. For large $\delta$, constant factors dominate. We chose the balancing strategy of bit insertions as it seemed most promising in preceding tests and reasonably analyzable. Yet there might be other, potentially more space-efficient strategies to be explored; see future work in Section 6.1.

# 6 Conclusion

With Consensus-CSF, we introduce a novel approach to compressed static functions and apply the Consensus technique to a less structured setting. For this, we utilize task insertions to balance task difficulties. We analyze our algorithm in detail by performing a ground-up analysis of Consensus' performance for less strict requirements on task difficulties than the original paper [17]. This is the main contribution of our paper. Our algorithm reaches arbitrarily close to the zeroth-order empirical entropy, reaching a space usage of $nH_0(1+\delta)$ for a construction time of $n \exp\left(\tilde{\mathcal{O}}\left(\sqrt{1/\delta}\right)\right)$ when assuming some properties of the value distribution to be constants.

We also implemented a slightly modified version of the algorithm. Our experiments show it reaching space usage mostly in the same order of magnitude as competitors for practical construction time, yet it never dominates any of them.

## 6.1 Future Work

With our implementation not surpassing competitors' performance, there are a couple of ideas to be explored to improve upon our approach. One idea is to optimize when to insert tasks using dynamic programming instead of the hard threshold used now to optimize a space usage–construction time tradeoff. Moreover, splitting heavy keys into multiple smaller keys using a Huffman code whose bits are stored individually in a CSF may improve performance for many heavy keys. Also, a filter could be applied to get rid of the most common values, making all other weights more balanced.

From a theoretical perspective, it might be interesting to analyze a version of the algorithm closer to the implementation, omitting the fallback data structure and allowing an arbitrary number of insertions. Further, completely different balancing schemes could be explored, like using a (weighted) k-perfect hash function.

Finally, CONSENSUS as analyzed in this paper could be applied to the bucket placement approach of MPHF construction to achieve faster query times.

## A Table of Variables

This section shows an overview of the variables used in this paper. The variables marked with $\star$ are the parameters that can be freely chosen in the implementation.

| | Variable | Value | Explanation |
|---|---|---|---|
| | $S$ | $\subseteq U$ | set of keys |
| | $U$ | | universe from which the keys come |
| | $n$ | $\lvert S\rvert$ | number of keys |
| | $\Sigma$ | | set of values |
| | $f$ | $: S \to \Sigma$ | function which we want to store |
| | $\rho_\alpha$ | $\lvert\{x \in S : f(x) = \alpha\}\rvert/n$ | frequency of value $\alpha \in \Sigma$ |
| | $\log_{1/2} \rho_\alpha$ | | weight of a key with value $\alpha$ |
| | $H_0$ | $\sum_{\alpha\in\Sigma} \rho_\alpha \log_{1/2} \rho_\alpha$ | zeroth order empirical entropy of the values $(f(x))_{x\in S}$ |
| | $I_{\max}$ | $\max_{\alpha\in\Sigma} \log_{1/2} \rho_\alpha$ | maximum information/weight of a key |
| | $I_{\min}$ | $\min_{\alpha\in\Sigma} \log_{1/2} \rho_\alpha$ | minimum information/weight of a key |
| | $\delta$ | $\in (0, \min_{\alpha\in\Sigma} \rho_\alpha^c)$ | tradeoff parameter |
| | $c$ | only implicitly defined | ensures $\log b/I_{\max} \geq c$ to simplify bounds |
| $\star$ | $b$ | $\frac{16}{\delta}$ | initial number of tasks per group |
| $\star$ | $\beta$ | $\sqrt{b} \log b$ | increment in which tasks are inserted |
| | $\epsilon$ | $\frac{\delta}{4}$ | bound on the loss if event $\mathcal{Q}_i$ occurs |
| | $a$ | $\log_{1/2} \epsilon$ | lower bound of $X_t$ for $\mathcal{Q}_i$ to occur |
| | $\lambda$ | $(b + \beta/2)(1 - \epsilon)/H_0$ | expected number of keys in one group |
| $\star$ | $\lambda H_0$ | | expected weight in one group |
| $\star$ | $D$ | | maximum difficulty inside a group; if there exists a task $t$ with $\log_{1/2} q_t > D$, the implementation inserts more tasks |
| $\star$ | $D_b$ | | maximum difficulty at the left border of a group; if the task $t = (i, 0)$ has difficulty $\log_{1/2} q_t > D_b$ the implementation inserts more tasks |
| | $i$ | $\in [n/\lambda]$ | group index |
| | $\ell_i$ | $\in \{0, 1, 2\}$ | number of insertions in group $i$ |
| | $m_i$ | $b + \ell_i\beta$ | number of tasks in group $i$ |
| | $j$ | $\in [m_i]$ | in-group task index |
| | $t$ | $(i, j)$ | task index, note $t + r := (i, j + r)$ |
| | $n/\lambda$ | | number of groups |
| | $\pi$ | $: U \to [n/\lambda]$ | partitions keys uniformly random into groups |

| Variable | Value | Explanation |
|---|---|---|
| $\pi_{i,\ell_i}$ | $: U \to [m_i]$ | partitions keys of group $i$ into tasks in that group. Randomly redrawn when $\ell_i$ changes. |
| $S_i$ | $\{x \in S : \pi(x) = i\}$ | keys in group $i$ |
| $W_i$ | $\sum_{x \in S_i} \log_{½} \rho_{f(x)}$ | weight of keys in group $i$ |
| $S_F$ | | keys in fallback |
| $W_F$ | $\sum_{x \in S_F} \log_{½} \rho_{f(x)}$ | weight of keys in fallback |
| $S_t$ | $\left\{x \in S_i : \pi_{i,\ell_i}(x) = j\right\}$ | keys in task $t = (i, j)$ |
| $W_t$ | $\sum_{x \in S_t} \log_{½} \rho_{f(x)}$ | weight in task $t$ |
| $W_{lr}$ | $\sum_{t=l}^{r-1} W_t$ | weight in the interval $[l, r)$ of tasks |
| $p_t$ | $\prod_{x \in S_t} \rho_{f(x)}$ | probability that a seed is successful for all keys in task $t$ |
| $q_t$ | $1 - (1 - p_t q_{t+1})^2$ | probability that Consensus will not backtrack from task $t$ |
| $\tilde{q}_t$ | $\tilde{q}_{i,m_i} = \min\left\{\tilde{q}_{i+1,0}, \epsilon 2^{-\beta}\right\}$ | $q_t$ with clamping at group borders |
| $X_t$ | $\log_{½} \tilde{q}_t$ | “difficulty” of task $t$ |
| $\text{loss}_t$ | $\log_{½}(1 - p_t \tilde{q}_{t+1}/2)$ | error term in linear approximation $X_t = X_{t+1} + W_t - 1 + \text{loss}_t$ |
| $\Delta_i$ | $X_{i,0} - X_{i,m_i}$ | change from the right to the left of group $i$ |
| $[l, r)$ | $\{l, \ldots, r - 1\}$ | interval of tasks inside a group |
| $\omega$ | $r - l$ | width of an interval $l, \ldots, r - 1$ of tasks |
| $\Delta_{lr}$ | $X_l - X_r$ | change over an interval of tasks $[l, r)$ |
| $\mathcal{G}_i$ | $\mathcal{X}_i \cap \mathcal{Q}_i$ | event that $i$ is a good group, no bumping |
| $\mathcal{B}_i$ | $\mathcal{G}_i^{\complement}$ | bad group, gets bumped |
| $\mathcal{X}_i$ | see Definition 3 | |
| $\mathcal{Q}_i$ | see Definition 3 | |
| $h_s$ | $: U \to \Sigma$ | seeded hash function s.t. for any seed $s$, key $x \in U$, and value $\alpha \in \Sigma$: $\Pr\left[h_{s(x)} = \alpha\right] = \rho_\alpha$. Implemented via alias table [27]. |
| $C$ | see Section 3 | Consensus vector |
| $I$ | see Section 3 | insertion vector, stores unary $\ell_i$ separated by zeros |
| $F$ | see Section 4.3 | fallback data structure |

## B Proofs

**Proof of Lemma 7.** We use the definition of the variables in Section 4.1 to prove the inequalities:

1. Since $\beta = \sqrt{b}\log b$ is monotone in $b$ and $b = \frac{16}{\delta} \geq 16$ we have $16 = \sqrt{16}\log 16 \leq \sqrt{b}\log b = \beta$. Let $u = \log b$. We have $u \geq 4$ since $b = \frac{16}{\delta} \geq 16$. Thus
$$\beta = \sqrt{b}\log b \leq b \Leftrightarrow u = \log b \leq \sqrt{b} = 2^{u/2}.$$
The last inequality holds, since at $u = 4$ we have $u = 2^{u/2}$, and for $u \geq 4$ the derivatives are $\frac{d}{du}u = 1$ and $\frac{d}{du}2^{u/2} = \frac{\ln(2)}{2}2^{u/2} \geq 2\ln(2) > 1$.
2. Since $\delta < 1$, $\epsilon = \frac{\delta}{4} < \frac{1}{4}$
3. Using that $\beta \leq b$ and $\epsilon < \frac{1}{4}$, we can show $\lambda H_0 = (1-\epsilon)\left(b + \frac{\beta}{2}\right) \geq \frac{3}{4}\frac{3}{2}\beta > \beta$ and $\lambda H_0 = (1-\epsilon)\left(b + \frac{\beta}{2}\right) \leq (b + \beta/2) \leq \frac{3}{2}b$.
4. We use the definitions of $\epsilon, b$ to show $b\epsilon = \frac{16}{\delta}\frac{\delta}{4} = 4$.
5. By definition $\lambda H_0/(1-\epsilon) = b + \beta/2$. Since $16 \leq \beta \leq b$ and $\epsilon b = 4$ we get $\epsilon\lambda H_0/(1-\epsilon) \leq \epsilon\frac{3}{2}b = 6 \leq 3/8\beta$.
6. We know $\log\delta \leq -cI_{\max}$ thus $\log\frac{1}{\delta} \geq cI_{\max}$. If we additionally use $b = \frac{16}{\delta}$ we get $\frac{\log b}{I_{\max}} \geq c\frac{\log\frac{16}{\delta}}{\log\frac{1}{\delta}} \geq c$.

◀

**Proof of Lemma 6.** Let $p := \tau\frac{\lambda}{n}$. First, we bound the moment generating function of $W$. Note that $W = \sum_{\alpha\in\Sigma}\log_{½}\rho_\alpha N^{(\alpha)}$ with $N^{(\alpha)} \sim \mathrm{Bin}(n\rho_\alpha, p)$ independent. For that we use that $x \mapsto e^{tx}$ is convex and therefore $e^{xt} \leq 1 + \frac{x}{I_{\max}}(e^{tI_{\max}} - 1)$ for $x \in [0, I_{\max}]$.

$$M_W(t) := \mathbb{E}[e^{Wt}] \overset{\text{indep.}}{=} \prod_{\alpha\in\Sigma}\mathbb{E}\left[\exp\left(N^{(\alpha)}\cdot t\log_{½}\rho_\alpha\right)\right] \overset{\text{Bin MGF}}{=} \prod_{\alpha\in\Sigma}\left(1 - p + pe^{t\log_{½}\rho_\alpha}\right)^{n\rho_\alpha}$$
$$\overset{\text{convex}}{\leq} \prod_{\alpha\in\Sigma}\left(1 - \not{p} + p\left(\not{1} + \frac{\log_{½}\rho_\alpha}{I_{\max}}\right)(e^{tI_{\max}} - 1)\right)^{n\rho_\alpha}$$
$$\overset{1+x\leq e^x}{\leq} \prod_{\alpha\in\Sigma}\left(\exp\left(p\frac{\log_{½}\rho_\alpha}{I_{\max}}(e^{tI_{\max}} - 1)\right)\right)^{n\rho_\alpha} = \exp\left(\frac{np}{I_{\max}}(e^{tI_{\max}} - 1)\underbrace{\sum_{\alpha\in\Sigma}\rho_\alpha\log_{½}\rho_\alpha}_{=H_0}\right)$$
$$= \exp\left(\tau\frac{H_0\lambda}{I_{\max}}(e^{tI_{\max}} - 1)\right).$$

1. Let $\substack{x>\mu \\ 0<x<\mu}$. We use a standard Chernoff bound:
$$\Pr\left[W \substack{\geq \\ \leq} x\right] \overset{\text{Chernoff}}{\leq} \inf_{t \substack{> \\ <} 0} M_W(t)e^{-tx} \leq \inf_{t \substack{> \\ <} 0}\exp\left(-tx + \frac{\mu}{I_{\max}}(e^{tI_{\max}} - 1)\right)$$
$$\overset{\star}{=} \exp\left(-\frac{1}{I_{\max}}x\ln\frac{x}{\mu} + \frac{\mu}{I_{\max}}\left(\frac{x}{\mu} - 1\right)\right) = \exp\left(-\frac{1}{I_{\max}}\left(x\ln\frac{x}{\mu} - x + \mu\right)\right)$$
For $\star$ we insert $t^* = \frac{1}{I_{\max}}\ln\frac{x}{\mu} \substack{> \\ <} 0$ where the minimum of $f(t) := -tx + \frac{\mu}{I_{\max}}(e^{tI_{\max}} - 1)$ is located, because
$$0 = f'(t) = -x + \mu e^{tI_{\max}} \Leftrightarrow t = \frac{1}{I_{\max}}\ln\frac{x}{\mu},$$
and $f''(t) = \mu I_{\max}e^{tI_{\max}} > 0$.
2. If we use the inequality $(1+u)\ln(1+u) - u \geq \frac{u^2}{2+\frac{2}{3}u} \quad (u > 0)$ [2:, Exercise 2.8], we get for $d \geq 0$

$$\Pr[W \geq \mu + d] \leq \exp\left(-\frac{1}{I_{\max}}\left((\mu+d)\ln\left(1+\frac{d}{\mu}\right) - (\mu+d) + \mu\right)\right)$$
$$= \exp\left(-\frac{\mu}{I_{\max}}\left(\left(1+\frac{d}{\mu}\right)\ln\left(1+\frac{d}{\mu}\right) - \frac{d}{\mu}\right)\right) \leq \exp\left(-\frac{\mu}{I_{\max}}\frac{(d/\mu)^2}{2+\frac{2}{3}\frac{d}{\mu}}\right)$$
$$= \exp\left(-\frac{1}{I_{\max}}\frac{d^2}{2\mu+\frac{2}{3}d}\right).$$

**3.** We use from [2: Exercise 2.8] the inequality for $u \in (0,1)$

$$-\ln(1-u) - u \leq \frac{u^2}{2(1-u)}$$
$$\overset{\cdot(u-1)}{\iff} (1-u)\ln(1-u) - u^2 + u \geq -\frac{u^2}{2} \overset{+u^2}{\iff} (1-u)\ln(1-u) + u \geq \frac{u^2}{2}$$

Thus, for $d \in (0,\mu)$

$$\Pr[W \leq \mu - d] \leq \exp\left(-\frac{1}{I_{\max}}\left((\mu-d)\ln\left(1-\frac{d}{\mu}\right) - (\mu-d) + \mu\right)\right)$$
$$= \exp\left(-\frac{1}{I_{\max}}\mu\left(\left(1-\frac{d}{\mu}\right)\ln\left(1-\frac{d}{\mu}\right) + \frac{d}{\mu}\right)\right)$$
$$\leq \exp\left(-\frac{1}{I_{\max}}\mu\left(\frac{(d/\mu)^2}{2}\right)\right) = \exp\left(-\frac{1}{I_{\max}}\frac{d^2}{2\mu}\right).$$

◀